\documentclass{article}

\usepackage{amssymb}
\usepackage{graphics}

\usepackage{amsmath}
\numberwithin{equation}{section}

\begin{document}
\renewcommand{\Bbb}{\mathbb}

\thispagestyle{empty}

%%%%%%%%%%%%%%%%%%%%%%%%%%%%%%%%%%%%%%%%%%%%%%%%%%%%%%%%%%%%%%%%%%%%%%%%%
%                            GREEK                                      %
%%%%%%%%%%%%%%%%%%%%%%%%%%%%%%%%%%%%%%%%%%%%%%%%%%%%%%%%%%%%%%%%%%%%%%%%%
\newcommand{\al}{\alpha}
\newcommand{\bet}{\beta}
\newcommand{\ga}{\gamma}
\newcommand{\del}{\delta}
\newcommand{\ep}{\epsilon}
\newcommand{\epx}{\varepsilon}
\newcommand{\ze}{\zeta}
\renewcommand{\th}{\theta}
\newcommand{\thx}{\vartheta}
\newcommand{\io}{\iota}
\newcommand{\la}{\lambda}
\newcommand{\ka}{\kappa}
\newcommand{\pix}{\varpi}
\newcommand{\rhx}{\varrho}
\newcommand{\si}{\sigma}
\newcommand{\six}{\varsigma}
\newcommand{\yp}{\upsilon}
\newcommand{\om}{\omega}
\newcommand{\phx}{\varphi}
\newcommand{\Ga}{\Gamma}
\newcommand{\De}{\Delta}
\newcommand{\Th}{\Theta}
\newcommand{\La}{\Lambda}
\newcommand{\Si}{\Sigma}
\newcommand{\Yp}{\Upsilon}
\newcommand{\Om}{\Omega}

%%%%%%%%%%%%%%%%%%%%%%%%%%%%%%%%%%%%%%%%%%%%%%%%%%%%%%%%%%%%%%%%%%%%%%%%%
%                         MATH MODE COMMAND                             %
%%%%%%%%%%%%%%%%%%%%%%%%%%%%%%%%%%%%%%%%%%%%%%%%%%%%%%%%%%%%%%%%%%%%%%%%%
\renewcommand{\L}{\cal{L}}
\newcommand{\M}{\cal{M}}
\newcommand{\G}{{\cal G}}
\newcommand{\J}{{\cal J}}
\newcommand{\Jb}{\bar{\cal J}}
\newcommand{\be}{\begin{equation}}
\newcommand{\ee}{\end{equation}}
\newcommand{\bea}{\begin{eqnarray}}
\newcommand{\eea}{\end{eqnarray}}
\newcommand{\jt}{\tilde{J}}
\newcommand{\Ra}{\Rightarrow}
\newcommand{\lra}{\longrightarrow}
\newcommand{\ti}{\tilde}
\newcommand{\pj}{\prod J}
\newcommand{\pjt}{\prod\tilde{J}}
\newcommand{\prb}{\prod b}
\newcommand{\prc}{\prod c}
\newcommand{\bft}{|\tilde{\phi}>}
\newcommand{\bfj}{|\phi>}
\newcommand{\lan}{\langle}
\newcommand{\ran}{\rangle}
\newcommand{\bz}{\bar{z}}
\newcommand{\bJ}{\bar{J}}
\newcommand{\tsp}{\tau\hspace{-1mm}+\hspace{-1mm}\si}
\newcommand{\tsm}{\tau\hspace{-1mm}-\hspace{-1mm}\si}
\newcommand{\vacr}{|0\rangle}
\newcommand{\vacl}{\langle 0|}
\newcommand{\IFF}{\Longleftrightarrow}
\newcommand{\phr}{|phys\ran}
\newcommand{\phl}{\lan phys|}
\newcommand{\non}{\nonumber\\}
\newcommand{\tg}{\tilde{g}}
\newcommand{\tM}{\ti{M}}
\newcommand{\hd}{\hat{d}}
\newcommand{\hL}{\hat{L}}
\newcommand{\sir}{\si^\rho}
\newcommand{\mf}{\mathfrak}
\newcommand{\mfg}{{\mathfrak{g}}}
\newcommand{\mfbg}{{\bar{\mathfrak{g}}}}
\newcommand{\mfk}{{\mathfrak{k}}}
\newcommand{\mfc}{{\mathfrak{c}}}
\newcommand{\mfbc}{{\bar{\mathfrak{c}}}}
\newcommand{\mfbk}{{\bar{\mathfrak{k}}}}
\newcommand{\mfn}{{\mathfrak{n}}}
\newcommand{\mfbn}{{\bar{\mathfrak{n}}}}
\newcommand{\mfh}{{\mathfrak{h}}}
\newcommand{\mfbh}{{\bar{\mathfrak{h}}}}
\newcommand{\mfp}{{\mathfrak{p}}}
\newcommand{\mfbp}{{\bar{\mathfrak{p}}}}
\newcommand{\mfU}{{\mf U}}
\newcommand{\p}{\partial}
\newcommand{\pb}{\bar{\hspace{0.2mm}\partial}}
\newcommand{\pl}{\partial_{+}}
\newcommand{\pmi}{\partial_{\hspace{-0.5mm}-}}
\newcommand{\Lg}{\L^{(\mfg})}
\newcommand{\Lgk}{\L^{(\mfg,\mfk)}}
\newcommand{\Lgkp}{\L^{(\mfg,\mfk')}}
\newcommand{\Lk}{\L^{(\mfk)}}
\newcommand{\Lkp}{\L^{(\mfk')}}
\newcommand{\Lc}{\L^{(\mfc)}}
\newcommand{\Lgg}{\L^{(\mfg,\mfg)}(\la)}
\newcommand{\Lgp}{\L^{(\mfg,\mfg')}(\la)}
\newcommand{\Mgp}{\M^{(\mfg,\mfg')}(\la)}
\newcommand{\Mgk}{\M^{(\mfg,\mfk)}(\la)}
\newcommand{\Hgk}{H^{(\mfg,\mfk)}}
\newcommand{\Mgkp}{\M^{(\mfg,\mfk^\prime)}}
\newcommand{\Hgp}{H^{(\mfg,\mfg')}}
\newcommand{\vo}{v_{0\la}}
\newcommand{\Ml}{\M(\la)}
\newcommand{\Ll}{\L(\la)}
\newcommand{\RR}{\mathbb{R}}
%%%%%%%%%%%%%%%%%%%%%%%%%%%%%%%%%%%%%%%%%%%%%%%%%%%%%%%%%%%%%%%%%%%%%%%%%
%                         MISCELLANEOUS                                 %
%%%%%%%%%%%%%%%%%%%%%%%%%%%%%%%%%%%%%%%%%%%%%%%%%%%%%%%%%%%%%%%%%%%%%%%%%

\newcommand{\e}[1]{\label{#1}\end{eqnarray}}
\newcommand{\bel}[1]{\be \label{#1}}
\newcommand{\rf}[1]{(\ref{#1})}
\newcommand{\ind}{\indent}
\newcommand{\noi}{\noindent}
\newcommand{\np}{\newpage}
\newcommand{\hs}{\hspace*}
\newcommand{\vs}{\vspace*}
\newcommand{\hsi}{\hspace*{1mm}}
\newcommand{\hsii}{\hspace*{2mm}}
\newcommand{\hsiii}{\hspace*{3mm}}
\newcommand{\hsv}{\hspace*{5mm}}
\newcommand{\vsi}{\vspace*{1mm}}
\newcommand{\vsii}{\vspace*{2mm}}
\newcommand{\vsiii}{\vspace*{3mm}}
\newcommand{\vsv}{\vspace*{5mm}}
\newcommand{\nl}{\newline}
\newcommand{\bqu}{\begin{quotation}}
\newcommand{\equ}{\end{quotation}}
\newcommand{\bit}{\begin{itemize}}
\newcommand{\eit}{\end{itemize}}
\newcommand{\ben}{\begin{enumerate}}
\newcommand{\een}{\end{enumerate}}
\newcommand{\ba}{\begin{array}}
\newcommand{\ea}{\end{array}}
\newcommand{\ul}{\underline}
\newcommand{\nn}{\nonumber}
\newcommand{\lef}{\left}
\newcommand{\rig}{\right}
\newcommand{\fra}{\twelvefrakh}
\newcommand{\Bb}{\twelvemsb}
\newcommand{\bT}{\bar{T}(\bz)}
\newcommand{\dagg}{^{\dagger}}
\newcommand{\qd}{\dot{q}}
\newcommand{\cP}{{\cal P}}
\newcommand{\hg}{\hat{g}}
\newcommand{\hh}{\hat{h}}
\newcommand{\hpg}{\hat{g}^\prime}
\newcommand{\htg}{\tilde{\hat{g}}^\prime}
\newcommand{\pri}{\prime}
\newcommand{\bis}{{\prime\prime}}
\newcommand{\lap}{\la^\prime}
\newcommand{\rhop}{\rho^\prime}
\newcommand{\Dgp}{\Delta_{g^\prime}^+}
\newcommand{\Dg}{\Delta_g^+}
\newcommand{\Pro}{\prod_{n=1}^\infty (1-q^n)}
\newcommand{\Pg}{P^+_{\hg}}
\newcommand{\Pgp}{P^+_{\hg\pri}}
\newcommand{\hmu}{\hat{\mu}}
\newcommand{\hnu}{\hat{\nu}}
\newcommand{\hrho}{\hat{\rho}}
\newcommand{\gp}{g^\prime}
\newcommand{\pp}{\prime\prime}
\newcommand{\CM}{\hat{C}(g',M')}
\newcommand{\CI}{\hat{C}(g',M^{\prime (1)})}
\newcommand{\CL}{\hat{C}(g',L')}
\newcommand{\HL}{\hat{H}^p (g',L')}
\newcommand{\HMI}{\hat{H}^{p+1}(g',M^{\prime (1)})}
\newcommand{\da}{\dagger}
\newcommand{\asu}{\ensuremath{\widehat{\hspace{0.5mm}\mathfrak{su}}(1,1)\,}}
\newcommand{\half}{\frac{1}{2}}
\newcommand{\wro}{w^\rho}
\newcommand{\sltwo}{\ensuremath{\widehat{\hspace{0.5mm}\mathfrak{sl}}_2} }
\renewcommand{\Box}{\rule{2mm}{2mm}}
\newcommand{\ads}{\ensuremath{\mbox{AdS}_{3}\,}}
\newcommand{\dd}{\ensuremath{D1\mbox{-}D5} }

%\begin{flushright}
%October 25, 2001 \\
%{\bf \sc Preliminary version}\\
%\end{flushright}
\vs{10mm}

\begin{center}

{\Large{\bf Sectors of solutions in Three-dimensional Gravity and Black Holes}}
\\
\vspace{10 mm}
{\large{Jens Fjelstad \footnote{email: jens.fjelstad@kau.se}\\ and\\
Stephen Hwang \footnote{email: stephen.hwang@kau.se}}
\vspace{4mm}\\
Department of Physics,\\ Karlstad
University, SE-651 88 Karlstad, Sweden}

\vs{15mm}

{\bf Abstract }\end{center} 
\begin{quotation}\noindent
We examine the connection between three dimensional gravity with negative cosmological constant and two-dimensional CFT via the Chern-Simons formulation. A set of generalized spectral flow transformations are shown to yield new sectors of solutions. One implication is that the microscopic calculation of the entropy of the Ba\~nados-Teitelboim-Zanelli (BTZ) black hole is corrected by a multiplicative factor with the result that it saturates the Bekenstein-Hawking expression.
\end{quotation}

\vs{6cm}
\begin{center}
October 2001\end{center}

\np
\setcounter{page}{1}

\section{Introduction}
\label{sec:introduction}
Three-dimensional gravity is a popular testing ground for ideas concerning gravity in four dimensions, in particular regarding questions about quantum gravity. Through the work of Townsend, Ach\'ucarro and Witten~\cite{Townsend:csgrav, Witten:gravcs} writing the Einstein-Hilbert action as a Chern-Simons (CS) term, there is a wide-spread belief that gravity in three dimensions is readily quantized. Indeed it seems a much more viable task than the realistic case, but there still remains a lot of work to actually complete the task of quantizing these Chern-Simons theories. One of the objectives of the present work is to investigate some properties of the quantum theory.
When formulated with a negative cosmological constant, this model is of particular interest.
One reason is the discovery by Ba\~nados, Teitelboim and Zanelli (BTZ)~\cite{BTZ} of a black hole which exists under these conditions, another reason is the AdS/CFT correspondence in three dimensions.

Although a substantial amount of work has been put into examining the quantum properties of the BTZ black hole, there is still a considerable gap in our understanding, in particular of the microscopic origin of the black hole entropy.
This issue is thoroughly discussed by Carlip~\cite{Carlip98}, where many references to different approaches and calculations may be found. Carlip argued convincingly that our present understanding gives no answer to questions like which degrees of freedom are responsible for the entropy, where (and indeed {\em if}) these degrees of freedom are localized, and most importantly whether they are abundant enough to account for the Bekenstein-Hawking entropy
\begin{equation}
   \label{bek-hawk}
   S = \frac{A}{4G}
\end{equation}
where $A$ is the horizon "area" and $G$ is Newton's constant.

In this paper we will show the presence of new sectors of solutions which will provide large enough degeneracy of states to account for the Bekenstein-Hawking entropy. These sectors are generated by generalized spectral flow transformations and correspond to gauge inequivalent solutions in the CS theory. Curiously enough, and in contrast to string theory on \ads, normal~\cite{Hwang:modinv, Maldacena:wzw1} spectral flow will just transform between canonically equivalent solutions. Therefore, although three-dimensional gravity can be described in terms of \sltwo currents, it does not seem to be equivalent to string theory on \ads.
A consequence of our result is that the question of where the degrees of freedom live becomes ill posed, rather the states correspond to certain classes of geometries. Moreover, as a consequence of our approach being completely gauge invariant, the results are independent of any particular choice of boundary conditions.

It should be emphasized that all calculations referred to above address these questions in pure gravity, i.e. there are no matter fields involved.
In most calculations of black hole thermodynamics, the presence of matter is crucial. For instance in the Euclidean path integral approach one really calculates thermodynamic properties of quantum fields in a static black hole background, and in string theory approaches one makes use of the full string theory partition function, which includes also matter fields. Therefore, it may appear unlikely that gravity alone would be enough to yield the Bekenstein-Hawking entropy. Still there are numerous claims that this is the case (see \cite{Carlip98} and references therein). These claims alone provide enough motivation for this study. Furthermore, regardless of the outcome, the study may shed considerable light on several aspects of quantum gravity.
The three dimensional setting of course simplifies matters considerably, and one may also on this ground question the relevance of the present study. This is a feature which is shared by all toy models in any area. There are, however, reasons to believe that this simple system is relevant in a broader sense. In string theory calculations, as pioneered by Strominger and Vafa~\cite{StrVafa}, the geometry of higher dimensional black holes are often in some limit a product of an \ads or BTZ geometry, and some other part. From the low energy effective field theory angle, three dimensional gravity (with negative cosmological constant) then plays a role also in these systems. One very interesting open problem is to relate the pure gravity calculations to the string theory calculations. Ultimately we would like to see how the gravity calculations, if correct, are embedded in the string approach.

Virtually all existing attempts rely on the relation of three-dimensional gravity to two-dimensional conformal field theory (CFT). This relation is motivated by Witten's result~\cite{Witten:qftjones} that Chern-Simons gauge theory is equivalent to chiral CFT, and also by the result of Brown and Henneaux~\cite{Brown-Henneaux} who showed that the asymptotic symmetry group of \ads is generated by a Virasoro algebra with a certain (classical) value of the central charge.
One approach due to Strominger~\cite{Strominger98} utilizes the Brown-Henneaux classical central charge 
\begin{equation}
   \label{bhcentralc}
   c = \frac{3l}{2G}
\end{equation}
together with a saddle point approximation of the partition function of the CFT to extract the asymptotic value of the density of states $\varrho(\Delta,\bar{\Delta})$. This is based on an old result by Cardy relating $\varrho$ to the central charge in a unitary modular invariant rational CFT.
Assuming that the underlying CFT is modular invariant, (\ref{bhcentralc}) can be inserted into the Cardy formula to yield the Bekenstein-Hawking expression for the BTZ black hole entropy.

As emphasized by Carlip, the central charge which appears in the Cardy formula is not the naive central charge, but rather the effective central charge
\begin{equation}
   \label{ceff}
   c_{eff} = c-24\Delta_{0}
\end{equation}
where $\Delta_{0}$ is the lowest $L_{0}$-eigenvalue. In a unitary rational CFT one has $\Delta_{0}=0$, but for theories such as Liouville theory or the $SL(2,\mathbb{R})$ WZNW model, this is no longer necessarily the case. Assuming that the CFT relevant for describing the BTZ black hole is the $SL(2,\mathbb{R})$ WZNW model, or some close relative such as Liouville theory, therefore makes the result of Strominger look odd from a purely gravitational perspective~\footnote{The existing partition functions for the $SL(2,\mathbb{R})$ WZNW model are formally divergent so it is a non-trivial task to extract physical information from them. It is, however, possible to extract a sensible regularized effective central charge, the value of which is $c_{eff}=3$}.

The nature of the degrees of freedom in Strominger's approach remains hidden since the only information used is the value of the central charge together with the assumptions of $\Delta_{0}=0$ and modular invariance. Also the question of where the degrees of freedom are located remains clouded, although it seems natural to associate them with the timelike boundary at asymptotic infinity. Other approaches, notably pioneered by Carlip~\cite{Carlip95}, treat the horizon as a boundary, and the relevant degrees of freedom are would-be-gauge modes which become physical at the boundary.

The paper is organized as follows.
In section two we recapitulate the gauge invariant connection between Chern-Simons theory and the chiral WZNW model, and we extend this to the Chern-Simons formulation of three-dimensional gravity. We also discuss unitarity of gravity in this formulation.
In section three we discuss some solutions, both well known and new, in the full gauge invariant theory. We move on to show in section four that the solutions are classified in $2|k|$ sectors corresponding to certain locality properties. Section five is devoted to the calculation of the entropy, and it is shown that the new classes of solutions provide exactly the correct amount of states to yield the Bekenstein-Hawking entropy by simply counting the number of states corresponding to a fixed asymptotic value of the mass and spin in the semiclassical limit. Also, some new light is shed on the original entropy calculation by Carlip~\cite{Carlip95}.
Finally various open questions are discussed, and also the relation to results in string theory. Since absolute normalizations are crucial, a list of conventions is included in the Appendix.

\section{The Chern-Simons formulation of Gravity}
\label{sec:csanalysis}
Our starting point is three-dimensional gravity defined through the Einstein-Hilbert action with negative cosmological constant
\begin{equation}
   \label{EHaction}
   I_{EH} = \frac{1}{16\pi G}\int_{\mathcal{M}}d^{3}x\sqrt{-g}\left( R-2\Lambda\right)
\end{equation}
where $\Lambda = -\frac{1}{l^{2}}$ is the cosmological constant.
Using
\begin{eqnarray}
   \label{CSconn1}
   A_{\al}^{\ a} & = & \omega_{\al}^{\ a} + \frac{1}{l}e_{\al}^{\ a}\\
   \label{CSconn2}
   A_{\al}^{'\ a} & = & \omega_{\al}^{\ a} - \frac{1}{l}e_{\al}^{\ a}
\end{eqnarray}
where $\omega_{\al}^{\ a}$ is the spin connection and $e_{\al}^{\ a}$ is the dreibein, we can rewrite (\ref{EHaction}) as~\cite{Townsend:csgrav, Witten:gravcs}
\begin{equation}
   \label{gravCS}
   I_{EH} = I_{CS}[A] - I_{CS}[A'] - I_{\partial}[A,A'].
\end{equation}
Here
\begin{equation}
   \label{CSaction}
   I_{CS}[A] = -\frac{k}{4\pi}\int_{\mathcal{M}}Tr\left[A\wedge dA + \frac{2}{3}A\wedge A\wedge A\right]
\end{equation}
defines a Chern-Simons action~\cite{Schonfeld, JackTemp, DesJackTemp1, DesJackTemp2} for a connection on a principal $SL(2,\mathbb{R})$-bundle and
\begin{equation}
    \label{EHbdry}
    I_{\partial} = \frac{k}{16\pi}\int_{\partial\mathcal{M}}Tr\left[ A\wedge A'\right]
\end{equation}
is a boundary term strongly resembling, although not equivalent to, the trace of the second fundamental form which is the conventional boundary term added to the Einstein-Hilbert action.
The parameters of the theories are related via
\begin{equation}
   \label{level}
   k = -\frac{l}{4G}.
\end{equation}
In the following analysis we will disregard the specific boundary term (\ref{EHbdry}) and instead take only the Chern-Simons term as the starting point for a canonical analysis of three-dimensional gravity. The motivation for this comes from our guiding principle being gauge invariance, and starting from only the bulk terms this principle will eventually lead to a unique boundary term in the Hamiltonian.
If we include the term (\ref{EHbdry}) in the canonical analysis, it looks almost impossible to achieve gauge invariance with less than canceling out the same term by adding a suitable boundary term.

The Chern-Simons gauge theory on a manifold with boundary is in itself connected to 
another theory, namely the WZNW model \cite{Witten:qftjones}. 
In a previous publication~\cite{Fjelstad-Hwang} we presented a formulation of this 
correspondence which is completely gauge invariant. Due to the gauge invariance the 
connection between the two theories is manifest and does not e.g. depend on any 
particular 
gauge choice or boundary condition. This will be of importance in our following 
analysis, 
as we will find general results which hold irrespectively of any particular boundary 
conditions. Let us first briefly recapitulate the basic ingredients.

The canonical analysis of the CS action (\ref{CSaction}) on the three-manifold 
$\mathcal{M}\cong \RR\times\Sigma$, where $\Sigma$ has the boundary $\partial\Sigma$, 
leads to the following primary and secondary constraints
\begin{equation}
\label{constraints}
P_a=0,\hs{10mm}\psi_a=\p_1A_{2a}-\p_2A_{1a}+f_{abc}A^b_1A^c_2=0.
\end{equation}
Here $P^a$ is the momentum conjugate to $A_{0a}$. By using the fundamental Poisson 
bracket\footnote{One should notice that the delta function defined through these 
brackets  satisfy $\int\delta^2(x-x')d^{2}x=1$ {\em even} for points $x'$ which lie on 
the boundary. Sometimes the delta function is defined so that its value on the 
boundary is half the one in the bulk. With this definition the fundamental Poisson 
brackets need to be adjusted in order to yield the correct equations of motion.} 
\begin{equation}
\{A_{2a}(x), A_{1b}(x')\}=-\frac{4\pi}{k}\eta_{ab}\delta^2(x-x')
\end{equation}
one easily shows that 
\begin{equation}
\label{eqcon}
\{G_\la,G_{\la^\prime}\}=G_{\la^{\prime\prime}}+\frac{4\pi}{k}\int_{\p \Sigma}dx^2(\p_2\la^a\la'_a+f_{abc}\la^a\la^{\prime b}A_2^c),\hspace{10mm} 
\la^{\prime\prime c}=\frac{4\pi}{k}f_{ab}^{~~c}\la^a\la^{\prime b}.
\end{equation}
Here $G_\la=\int_{\Sigma}d^2x(\p_1A_{2a}-\p_2A_{1a}+f_{abc}A^b_1A^c_2)\la^a$ and we 
have chosen $x^2$ to parameterize $\p \Sigma$. Equation (\ref{eqcon}) implies that the secondary constraints $\psi_a=0$ are second class. Gauge invariance is broken by the boundary terms. To make the theory gauge invariant we proceed in two steps. First we modify the action by adding a boundary term of the form $\int_{\M} A_{0}^aA^2_a$. Then we introduce new degrees of freedom on $\p\M$, namely the currents $J_{a}(x^0,x^2)$ satisfying the equal time algebra
\begin{equation}
\label{sl2}
\{J_a(x^2), J_b(x^{\prime 2})\}=-f_{ab}^{\
   \, c}J_{c}\del (x^2-x^{\prime 2})-\frac{k}{4\pi}\eta_{ab}\partial_{2}\del(x^2-x^{\prime
   2}).
\end{equation}
Finally we define a new Hamiltonian
\begin{equation}
H={k\over 4\pi}\int_\Sigma d^2x\left(\p_{1}A_{2a}-\p_{2}A_{1a} + 
f_{abc}A_{1}^bA_{2}^c\right)A_{0}^a+{k\over 4\pi}\int_{\p\Sigma}dx^2\left( 
A_{2a}|_{\p \Sigma}+\frac{4\pi}{k}J_{a}\right)A^{a}_0|_{\p \Sigma}
\end{equation}
which yields the following secondary constraints 
\begin{equation}
\label{1stclconstr}
\psi'_a=\p_{1}A_{2a}-\p_{2}A_{1a} + f_{abc}A_{1}^bA_{2}^c+(A_{2\, 
a}+\frac{4\pi}{k}J_{a})\delta(x\in\p\Sigma)=0.
\end{equation}
They satisfy a first class algebra 
$\{G'_\la,G'_{\la^\prime}\}=G'_{\la^{\prime\prime}}$ for $G'_\la=\int\psi'_a\la^a$ 
and $\la^{\pp}$ as in (\ref{eqcon}). The delta function 
$\delta(x\in\p\Sigma)$ is defined by 
$\int_{\Sigma}d^2xf(x)\delta(x\in\p\Sigma)=\int_{\p\Sigma}[f(x)]_{\p\Sigma}$. Notice 
that the currents $J^a$ enter into the Hamiltonian as  line source terms, showing 
that the presence of the boundary "induces", for consistency, a source at the 
boundary. Notice also that if the boundary consists of separate disjoint pieces, each 
piece will need a separate source like term, which in each case is represented by a 
current $J^a$. 

As was shown in~\cite{Fjelstad-Hwang}, the modification due to the current $J_a(x^2)$ essentially does 
not change the theory. By fixing the gauge on the boundary the currents may be 
eliminated leaving the original theory (with a particular boundary term 
added), in which gauge invariance is broken due to the 
boundary. 

The boundary term in (\ref{1stclconstr}) means that for smooth vector fields 
$A_{i}^{\ 
a}$, i.e. neither the fields nor their derivatives contain any delta function 
singularities, the constraints imply
\begin{equation}
   \label{JtoA}
   J_{a} = -\frac{k}{4\pi}A_{2\, a}|_{\partial\Sigma}.
\end{equation}
This relation was first found in \cite{Banados95}. 
If we would not have chosen $x^2$ to parameterize the boundary, then this identification will be replaced by\be
 J_{a}(\phi) = \frac{k}{4\pi}A_{t\, a}(x(\phi)).
 \end{equation}
 where $A_t$ is the tangential component of $A$ relative to the boundary and $\phi$ 
 parameterizes the boundary.  One may exhibit the \sltwo symmetry hidden in the constraints even 
 more directly. Define
\begin{equation}
\label{Jtilde}
\tilde{J}_a(x^2)=\frac{k}{4\pi}\int dx^1\left(\p_{1}A_{2a}-\p_{2}A_{1a} + f_{abc}A_{1}^bA_{2}^c\right)-\frac{k}{4\pi}A_{2\,a}(x)|_{\partial\Sigma}. 
\end{equation}
$\tilde{J}_a(x^2)$ generate an \sltwo current algebra of opposite 
level to $J_a(x^2)$, so that $J_a(x^2)+\tilde{J}_a(x^2)$ gives a
centerless algebra. By (\ref{1stclconstr}) we find that $J_a+\tilde{J}_a=0$ on the 
constraint surface.

The constraints (\ref{1stclconstr}) contain also Virasoro constraints. Define
\bel{invariantT}
T(x^2)=\left[-{2\pi\over k}(J^a-\tilde{J}^a) + 
\beta^a\partial_2\right](J_a+\tilde{J}_a)=-{2\pi\over 
k}\left[(J)^2-(\tilde{J})^2\right]+ \beta^a\partial_2(J_a+\tilde{J}_a),
\end{equation}
where $\beta^a$ are constants. $T(x^2)$ is a twisted Sugawara stress energy tensor. 
If $\p\Sigma$ is a circle, the Fourier modes of
$T(x^2)$, which are the Virasoro generators, satisfy a Virasoro algebra where  the 
central 
charge $c$ is zero at the classical level. As the Virasoro generators vanish on the 
constraint surface we conclude that they generate gauge transformations
on $\mathcal{M}$, including the boundary. Reparametrizations on the boundary are, 
therefore, included as a subset of the original gauge transformations. 

Let us now show that although there exist Virasoro constraints on the boundary, they 
will not be able to eliminate all time-like excitations. The basic reason is that the 
reparametrizations of the WZNW model are coupled to the transformations of the vector 
field and both of them include time-like degrees of freedom. 
We use the gauge invariance on all of $\mathcal{M}$, including the boundary, to 
impose the gauge 
\begin{equation}
\label{gauge}
A_{0}^{\ a}=0, \hsv A_{1}^{\ a}=0.
\end{equation}
The first condition fixes 
the trivial constraint $P_a=0$. Let us now check that the second condition fixes the gauge 
completely. This is the case if the
equation
\begin{equation}
\label{A2fix}
\{G'_{\lambda}, A_{1a}\}=0
\end{equation}
has the unique solution $\la_a(x)=0$ on the constraint surface. This equation  
is equivalent to $\partial_1\la_a(x)-\la_a(x)\del(x\in\partial{\mathcal{M}})=0$ on the constraint surface. 
As we require the gauge parameter and its derivative to be smooth, this equation has 
the 
unique solution $\la_{a}(x)=0$, as required. We may now solve the constraints. From 
(\ref{1stclconstr}) we get for smooth vector fields that $\partial_1A^{\, 
a}_{2}(x)=0$ and
$
A_{2}^{\, a}(x)|_{x\in\partial\Sigma}=-\frac{4\pi}{k}J^a(x^2)$, which implies 
\begin{equation}
\label{solveingauge}
A_{2}^{\, a}(x)=-\frac{4\pi}{k}J^a(x^2).
\end{equation}
This solves all constraints and the physical subspace is spanned by $J^a(x^2)$. As 
these \sltwo currents are completely unconstrained, we still have a time-like 
degree of freedom in $J^0(x^2)$. In quantizing this theory these degrees of freedom will, 
in general, result in a breakdown of unitarity. 

It is now possible to write a gauge-fixed Hamiltonian for this new theory. An  elegant way 
is to introduce the BRST charge for the extended theory in the BFV formalism~\cite{
Fradkin:bfv1,Batalin:bfv2}. This was done in~\cite{Fjelstad-Hwang}, where 
different gauge fixing fermions were shown to yield the CS and WZNW theories, 
respectively. 
The gauge fixing fermion
\begin{equation}
   \label{wzwgfix}
   \chi= \int_{\Sigma} \bar{c}_{a}\dot{A}^{\ a}_{0} + \int_{\p\Sigma} \bar{c}_{a}(J^{a}-\frac{\alpha kl}{2}P^{a})
\end{equation}
corresponds to the chiral WZNW model, and the corresponding total Hamiltonian is
\begin{equation}
   \label{gaugefixedhamiltonian}
   H =\int_\Sigma dx^2(P^a\dot{A}_a^0+\bar{b}^a\dot{\bar{c}}_a)+
  \int_{\partial\Sigma}dx^2 \left[ P^{a}(J_{a}-\frac{\alpha kl}{2}P_a)+f_{ab}^{~~ c}J_{c}\bar{c}^{b}c^{b}\right] 
\end{equation}
with $\al$ an arbitrary parameter. Here $A^{\ a}_{0}$, $P_a$ and the ghosts $\bar{b}_a$, 
$\bar{c}_a$ are non-dynamical. Elimination of these by the equations of motion yields the standard chiral Sugawara Hamiltonian 
\begin{equation}
\label{CSHam}
H=\frac{1}{2\al kl}\int_{\partial \Sigma}dx^2J^aJ_a.
\end{equation}

At this point we should again make contact with 3d gravity.  From  (\ref{gravCS}) it is equivalent to the difference of two CS theories, or more exactly to the sum of a CS theory 
with level $k$ and another of level $-k$. By 
(\ref{CSHam}) we generally have a corresponding Hamiltonian
\begin{equation}
\label{Hamgen}
H=\frac{1}{2\al kl}\int dx^2 J^aJ_a+\frac{1}{2\al'kl}\int dx^2 J^{\prime a}J'_a,
\end{equation}
where $J_a$ has level $k$ and $J'_a$ has level $-k$. Due to the factors of $l$ 
included, the constants $\al$ and $ \al'$ are dimensionless constants that are 
taken to be independent of $k$. Depending on the values of $\al$ and $\al'$ this 
Hamiltonian corresponds two distinctly different cases, namely a chiral - 
anti chiral theory ($\al\neq -\al'$) or a chiral-chiral theory ($\al= -\al'$). This is 
seen as follows. The equations of 
motion following from (\ref{Hamgen}) are
\begin{equation}
\p_0J_a=-{1 \over 4\pi\al l}\p_2J_a, \hs{10mm}\p_0J'_a={1 \over 4\pi\al'l}\p_2J'_a.
\end{equation}
Defining new coordinates $x^+=\half(x^0-{4\pi\al l}x^2)$ and 
$x^-=\half(x^0+{4\pi\al'l}x^2)$, which are independent provided $\al\neq -\al'$, the 
equations of motion simply read $\p_-J_a=0$ and $\p_+J'_a=0$, showing that we have a 
non-chiral theory. We can in this case without loss of generality fix $\al=\al'$. For 
$\al= -\al'$ we instead take $x^\pm=\half(x^0\mp{4\pi\al l}x^2)$ and we have 
$\p_+J_a=\p_+J'_a=0$ yielding a chiral-chiral theory. The Hamiltonians read
\begin{equation}
\label{nonchiralH}
H_{\tt nonchir}=\frac{1}{2\al kl}\int dx^2 (J^a(x^+)J_a(x^+)+J^{\prime a}(x^-)J'_a(x^-))
\end{equation}
\begin{equation}
\label{2chiralH}
H_{\tt chir-chir}=\frac{1}{2\al kl}\int dx^2 (J^a(x^-)J_a(x^-)-J^{\prime a}(x^-)J'_a(x^-))
\end{equation}
One should emphasize that both cases contain exactly the same degrees of freedom, namely two $\sltwo$ currents of opposite levels and they are classically 
equivalent (but not at the quantum level). To see which Hamiltonian is relevant for our case, one must go back to the starting point, the Einstein-Hilbert action 
(\ref{EHaction}). This action is trivially invariant under the change $l\rightarrow -l$, as the cosmological constant $\La=-1/l^2$. By (\ref{JtoA}), (\ref{CSconn1}) and 
(\ref{CSconn2}) this translates into
\begin{equation}
\label{l-transformation}
J\rightarrow J', \hs{10mm}J'\rightarrow J\hs{10mm}k\rightarrow -k.\vs{3mm}
\end{equation}
Using (\ref{JtoA}) we must therefore require our Hamiltonian to be 
invariant under (\ref{l-transformation}). This is only true for $H_{\tt nonchir}$  as
$\al$ is invariant under the transformation, so our theory is non-chiral. 

As remarked already in \cite{Witten:gravcs} one may 
modify the Einstein-Hilbert theory by adding the Hamiltonian (\ref{2chiralH}) (multiplied 
by an arbitrary factor), without changing the classical theory. As we 
have just seen this will still, in general, with suitably modified coordinates, give a non-chiral theory. 

The Hamiltonians above are not completely gauge-fixed. There still remains 
an invariance under global $SL(2,\RR)$ transformations. Classically one could 
fix all the values $J^a_0$, $a=1,2,3$, i.e. the zero modes in the 
Fourier decomposition of $J^a(x^2)$ when $x^2$ is an angular coordinate, but in preparation for our quantum treatment we only fix the Cartan generator $J^0_0$. 
Requiring this constant to take values in a unitary representation of $SL(2,\RR)$ 
requires us to fix $J^0_0$ to a value
\begin{equation}
\label{unitaryrepr}
J^0_0=a+m,
\end{equation}
where $m\in\mathbb{Z}$ and $a=0$ or $1/2$. The representations of discrete highest or 
lowest weight have a highest or lowest value of $m$, while for the continuous 
representations all values are present. The representations are characterized by the 
value of the quadratic Casimir $C^{(2)}$ and for the continuous representations also 
by $a$. In our conventions we have $C^{(2)}=\eta_{ab}J^a_0J^b_0$ and take values 
$-j(j+1)$, where $2j$ is real and integer for the discrete representations and 
$j=-1/2+i\rho$, $\rho\in\mathbb{R}$ for the principal continuous representations. As 
we will thoroughly discuss in section~\ref{sec:sectors}, different values in 
(\ref{unitaryrepr}) will be related to different sectors of solutions.

\section{Known and new solutions}
\label{sec:solutions}
We will be interested in solutions $A_{\mu a}$ of the  bulk constraints 
\begin{equation}
\label{bulkconstraints}
\p_{1}A_{2a}-\p_{2}A_{1a} + f_{abc}A_{1}^bA_{2}^c=0.
\end{equation}
These solutions correspond to vacuum solutions of the theory i.e. {\em outside all 
sources}. This may seem as a contradiction, as the WZNW currents look like sources 
and the 
boundaries are certainly part of the manifold. However, as we would like to stress, 
the currents enter as source terms, {\em but they will not enforce any source-like 
singularities at the boundaries}. Instead, we will always take our gauge fields to be smooth near and on the boundary and the currents will, therefore, only enforce the identification (\ref{JtoA}). Apart from the requirement that our fields are smooth on the manifold, there are no restrictions on the boundary. Hence, we have a large amount of freedom in the choice of what we call the boundary of our manifold.

The solution to eq.(\ref{bulkconstraints}), and the corresponding one for $A^{\prime a}_\mu$, which is of particular interest is the BTZ black hole solution
\bea
   A^{0} & = & N_{\bot}(\frac{1}{l}dt+d\phi )\nonumber\\
   \label{btzconnection1}
   A^{1} & = & N_{\bot}^{-1}\left(\frac{1}{l} - N_{\phi}\right)dr\\
   A^{2} & = & \left(rN_{\phi}+\frac{r}{l}\right) \left(\frac{1}{l}dt+d\phi\right)\nonumber
\eea
\bea
   A^{\prime 0} & = & -N_{\bot}(\frac{1}{l}dt-d\phi )\nonumber\\
    \label{btzconnection2}
   A^{\prime 1} & = & -N_{\bot}^{-1}\left(\frac{1}{l} + N_{\phi}\right)dr\\
   A^{\prime 2} & = & \left(-rN_{\phi}+\frac{r}{l}\right) \left(\frac{1}{l}dt-d\phi\right).\nonumber
\eea
which gives the BTZ metric \cite{BTZ}
\be
   \label{btzmetric}
   ds^{2} = -N_{\bot}^{2} dt^{2} + N_{\bot}^{-2} dr^{2} + r^{2}\left(d\phi + N_{\phi} dt\right)^{2}.
\end{equation}
The lapse and shift functions, $N_{\bot}$ and $N_{\phi}$, are given by
\bea
   \label{lapse}
   N_{\bot} & = & \frac{1}{rl}(r^2-r_+^2)^\half (r^2-r_-^2)^\half\\
   \label{shift}
   N_{\phi} & = & -\frac{r_+r_-}{r^{2}l}
\eea
The inner and outer horizons, $r_{-}$ and $r_{+}$, of the black hole are related to the mass $M$ and angular momentum $J$ via 
\begin{equation}
   \label{inoutradii}
   M=\frac{r_{+}^{2}+r_{-}^{2}}{8Gl^{2}} \quad \mbox{and} \quad J=\frac{r_{+}r_{-}}{4Gl}.
\end{equation}
This metric describes a black hole only if $|J|\leq Ml$, otherwise there is generically a naked conical singularity at $r=0$. Although the parameter $M$ 
describes the mass, it still makes sense to continue it to values below zero. The interval $-1<8GM<0$ corresponds to a naked conical singularity at $r=0$, but for the 
value $8GM=-1$ the deficit angle is zero and the metric describes (the infinite cover of) AdS$_{3}$. The solution is not real when $r_-<r<r_+$ i.e. when $N_\perp$ is imaginary. However, it is still straightforward to construct the corresponding solutions for this case.
 
The relation (\ref{JtoA}) gives us classically the $\widehat{\hspace{0.5mm}\mathfrak{sl}}_2$ currents in the gauge fixed version 
corresponding to the WZNW model. If we take a boundary at $r=R$ parameterized by $x^2=\phi$ then, since the solution is $\phi$-independent, this relation reads 
\begin{equation}
\label{JtoAa}
J^a_0=-\frac{k}{2}A_2^{\ a}|_{\p \Si},\hsv J^{\prime a}_0=-\frac{k}{2}A_2^{\prime a}|_{\p \Si}
\end{equation}
where $J^a_0, J^{\prime a}_0$ are the Fourier zero modes of the currents. In particular, for the solution (\ref{btzconnection1}) and (\ref{btzconnection2})
\begin{equation}
J^0_0=J^{\prime 0}_0=-\frac{k}{2}\left| N_\perp(R)\right|.
\end{equation}
These constant values are valid for all $R>0$.

There exist more interesting solutions of the same type. It was recently 
shown~\cite{Sundborg:multiadsbh} that for each choice of the BTZ parameters M and J 
there exist infinitely many gauge inequivalent geometries which all 
have asymptotically the same BTZ geometry. These geometries correspond generically to 
multi-centered black hole solutions, where the centers refer to the 
singularities. The singularities in question may be hidden behind the same horizon, 
or they may have separate 
horizons justifying the name multi black hole solutions. In one set of coordinates (cf.~\cite{Sundborg:multiadsbh}), 
the BTZ metric outside the outer horizon can be written
\begin{equation}
   \label{BTZmetric2}
   ds^{2}=-\sinh^{2}(\rho-\alpha-\frac{\pi}{2})[r_{+}dt- 
   r_{-}d\phi]^{2}+d\rho^{2}+\cosh^{2}(\rho-\alpha-\frac{\pi}{2})[r_{-}dt-r_{+}d\phi]
   ^{2}
\end{equation}
where $\alpha=\arctan\left(\frac{r_{-}}{r_{+}}\right)$. In the CS formulation one choice of gauge connections corresponding to this metric is
\bea
   A^{0} & = & -\frac{1}{l}(r_{+}-r_{-})\sinh(\rho-\alpha-\frac{\pi}{2})\left(\frac{1}{l}dt + d\phi\right)\nonumber\\
   \label{btzconnection3}
   A^{1} & = & \frac{1}{l}(r_{+}-r_{-})\cosh(\rho-\alpha-\frac{\pi}{2})\left(\frac{1}{l}dt+ d\phi\right)\\
   A^{2} & = & \frac{1}{l}d\rho\nonumber
\eea
\bea
   A^{'0} & = & -\frac{1}{l}(r_{+}+r_{-})\sinh(\rho-\alpha-\frac{\pi}{2})\left(\frac{1}{l}dt-d\phi\right)\nonumber\\
   \label{btzconnection4}
   A^{'1} & = & \frac{1}{l}(r_{+}+r_{-})\cosh(\rho-\alpha-\frac{\pi}{2})\left(\frac{1}{l}dt-d\phi\right)\\
   A^{'2} & = & -\frac{1}{l}d\rho.\nonumber
\eea
For the multi-centered solutions this is generalized to
\bea
   A^{0} & = & -g_{0}(h)\left(\frac{1}{l}dt+f\right)\nonumber\\
   \label{multicbh}
   A^{1} & = & g_{1}(h)\left(\frac{1}{l}dt+f\right)\\
   A^{2} & = & dh\nonumber
\eea
with similar expressions for $A'$.
Here $h$ is a parameter generalizing the coordinate $\rho$ and $f$ is a spatial one-form which is 
closed except in isolated points (the singularities)
\begin{equation}
   \label{clform}
   df = 2\pi \sum_{i}q_{i}\del^{(2)}(\vec{x}-\vec{x}_{i}).
\end{equation}
The functions $g_{0}$ and $g_{1}$ satisfy
\begin{equation}
   \label{hypeqns}
   \frac{dg_{0}}{dh}=-g_{1} \qquad \frac{dg_{1}}{dh}=-g_{0}
\end{equation}
which means that they are hyperbolic functions. Demanding that these solutions are asymptotically BTZ yields that (at least for large $h$) $g_{0}=\sinh(h)$ and $g_{1}=\cosh(h)$.
The asymptotic behavior of $f$ is also determined. For large $h$ we must have $f=Qd\phi$ for some constant $Q$, the meaning of which is clear for the BTZ expression. In that case 
we have $lQ=r_{+}-r_{-}$. For an $n$-center solution we have $Q=\sum_{I=1}^{n}q_{i}$ where $q_{i}$ are charges in the CS theory measuring the strength of the sources. In analogy 
with the single black hole we can write $lq_{i}=r_{i+}-r_{i-}$ which yields
\begin{equation}
   \label{sumrule}
   r_{+}-r_{-}=\sum_{i=1}^{n}(r_{i+}-r_{i-}).
\end{equation}

Solutions to 
the equations (\ref{bulkconstraints}) can be found in terms of 
Wilson loops. Being gauge invariant in the bulk, i.e. up to boundary corrections, they are relevant to 
use as tools for exploiting these issues. The Wilson loop is defined by
\begin{equation}
W[{\cal C}]= P\left\{e^{-\oint_{\cal C} A}\right\}
\end{equation}
where $P$ stands for path ordering. 

For the BTZ black hole it was shown in \cite{Cangemi-L-M} that the trace of the 
Wilson loop may be expressed solely in terms of the Casimirs of the $so(2,2)\sim 
sl(2,\RR)\oplus sl(2,\RR)$ algebra. One finds for a Wilson loop encircling the origin
\begin{equation}
\label{BTZWilson}
Tr(W[{\cal C}])= Tr(e^{{2\pi\over l}(r_+-r_-)T_2})=2\cosh [{\pi(r_+-r_-)\over l}].
\end{equation}
For the other sector described by $A'$ one has similarly
\begin{equation}
\label{BTZWilson1}
Tr(W'[{\cal C}])= Tr(P\left\{e^{-\oint_{\cal C} A'}\right\})=Tr(e^{{{2\pi\over l}(r_+-r_-)}T_2})=2\cosh [{\pi(r_++r_-)\over l}].
\end{equation}

By using the BTZ solution (\ref{btzconnection1}) and (\ref{btzconnection2}) and the correspondence (\ref{JtoA}) one finds classically the Casimirs 
\bea
   C^{(2)} & = & \eta_{ab}(\int_0^{2\pi}J^{a} d\phi )(\int_0^{2\pi} J^{b}d\phi ) \nonumber \\
   \label{btzreps}
   & = & 2Gk^{2}\left(M-\frac{J}{l}\right)=\left({k(r_+-r_-)\over 2l}\right)^2.
\eea
and
\bea
   C^{ (2)\prime} & = & \eta_{ab}(\int_0^{2\pi}J^{\prime a} d\phi )(\int_0^{2\pi} J^{\prime b}d\phi  )\nonumber \\
   \label{btzreps3}
   & = & 2Gk^{2}\left(M+\frac{J}{l}\right)=\left({k(r_++r_-)\over 2l}\right)^2.
\eea
Then
\begin{equation}
\label{BTZWilson2}
Tr(W[{\cal C}])=2\cosh [{2\pi \sqrt{C^{(2)}}\over |k|}],\hsv Tr(W'[{\cal C}])=2\cosh [{2\pi \sqrt{{C}^{(2)\prime}}\over |k|}].
\end{equation}
For the multi-black hole solutions one gets a similar expression \cite{Sundborg:multiadsbh}, where in (\ref{btzreps}) and (\ref{btzreps3}) ($r_+- r_-$) is given by (\ref{sumrule}) and an analogous expression for ($r_++r_-$).

It is simple to get further solutions to the constraints (\ref{bulkconstraints}) in analogy with the multi-centered solutions. In order to emphasize the vast amount of gauge inequivalent solutions which are asymptotically BTZ we will show an example of one such generalization, namely a line-source. All solutions described so far correspond in the CS description to point-like sources, but it is equally simple to describe 1-dimensional charge distributions, i.e. line-sources~\footnote{We hesitate somewhat to call these solutions black strings since their relation to the existing three-dimensional black string solutions is unclear.}.
A convenient choice of spatial one-form $f(\vec{x})$ describing a point-charge at the point $(x_{0},y_{0})$ is provided by $G(x,y;x_{0},y_{0})$
\begin{equation}
   \label{CSGreensfn}
   G(x,y;x_{0}y_{0}) = q\frac{(x-x_{0})dy - (y-y_{0})dx}{(x-x_{0})^{2} + (y-y_{0})^{2}}.
\end{equation}
Viewed as a Greens function for the exterior derivative we can construct a line-source supported on the curve $\gamma$ through
\begin{equation}
   \label{CSlinef}
   f^{[\gamma]}(x,y) = \int_{\gamma}\sigma(x',y')G(x,y;x',y')ds'
\end{equation}
where $\sigma$ is the line-source density (i.e. the total charge $Q$ is $\sigma$ integrated over the curve $\gamma$), and the one-form $f^{[\gamma]}(x,y)$ is used exactly as before.

As a simple example we construct a straight line-source on the interval $[-a,a]$ on the $x$-axis, and with total charge $Q$. A full solution also includes $A'$, but for simplicity we will write down only $A$. The other part is extracted in full analogy with the ordinary BTZ solution.
\bea
   f^{[-a,a]}(x,y) & = & \frac{Q}{2a}\int_{-a}^{a}\frac{(x-x')dy-ydx}{(x-x')^{2}+y^{2}}dx'\nonumber\\
   \label{straightlinef}
   & = & \frac{Q}{2a}\left[ \frac{1}{2}\ln{\left| 1-\frac{4ax}{(x+a)^{2}+y^{2}}\right|}dy - \arctan{\left(\frac{2ay}{x^{2}+y^{2}-a^{2}}\right)}dx\right]\\
   & = & \frac{Q}{2a}b(x,y).\nonumber
\eea
To leading order in $1/\rho$ where $\rho^{2} = x^{2}+y^{2}$ we find
\begin{equation}
   \label{fasympt}
   f^{[-a,a]}(\rho,\phi) \rightarrow Qd\phi
\end{equation}
which inserted in (\ref{multicbh}) gives $A$ asymptotically, and analogously for $A'$
\begin{equation}
   \label{fprimeasym}
   f^{'\ [-a,a]}(\rho,\phi) \rightarrow Q'd\phi.
\end{equation}
A suitable choice of the function $h$ then yields an asymptotically BTZ metric also for line-sources.
For generic CS configurations one expects additional degeneracies of the metric, this was indeed the reason for restricting to a narrow class of solutions in~\cite{Sundborg:multiadsbh}. It is our firm belief, however, that to consistently quantize gravity we must include all of these solutions, and there is no obvious reason why the "black lines" would contribute any less to the space of states than the multi-center black hole solutions.
To examine the degeneracies of this solution we use the relations (\ref{CSconn1}) and (\ref{CSconn2}) to extract the dreibein $e = \frac{1}{2}(A-A')$, and we obtain the metric
\bea
   ds^{2} & = & 2Tr(ee)\nonumber\\
   \label{linemetric}
   & = & \lambda_{1}(h)dt^{2} + \frac{1}{4a^{2}}\lambda_{2}(h)b^{2}(x,y) + \frac{Q_{+}Q_{-}}{a}b(x,y)dt + dh^{2}
\eea
where we have introduced $Q_{\pm} = \frac{1}{2}(Q\pm Q')$ and
\bea
   \label{lambda1}
   \lambda_{1}(h) & = & -Q_{-}^{2}g_{0}^{2}(h) + Q_{+}^{2}g_{2}^{2}(h)\\
   \label{lambda2}
   \lambda_{1}(h) & = & -Q_{+}^{2}g_{0}^{2}(h) + Q_{-}^{2}g_{2}^{2}(h)
\eea
and we have assumed that $-g_{0}^{2}(h) +g_{2}^{2}(h) = 1$.
Writing $b(x,y) = A(x,y)dx + B(x,y)dy$ the metric components are
\begin{equation}
   \label{lmet}
   g_{\al\bet} = \left[ \begin{array}{ccc}
   			   \vs{2mm}
   			   \lambda_{1} & \frac{Q_{+}Q_{-}}{2a}A & \frac{Q_{+}Q_{-}}{2a}B\\
			   \vs{2mm}
   			   \frac{Q_{+}Q_{-}}{2a}A & \lambda_{2}\hspace{-1mm}\left(\frac{A}{2a}\right)^{2} + \left(\partial_{x}h\right)^{2} & \frac{\lambda_{2}}{4a^{2}}AB + \partial_{x}h \partial_{y}h\\
   			   \frac{Q_{+}Q_{-}}{2a}B & \frac{\lambda_{2}}{4a^{2}}AB + \partial_{x}h \partial_{y}h & \lambda_{2} \hspace{-1mm}\left(\frac{B}{2a}\right)^{2} + \left(\partial_{y}h\right)^{2}
   			\end{array}
   			\right]
\end{equation}
with determinant
\begin{equation}
   \label{linedet}
   g = \frac{1}{4a^{2}}\left[B(x,y)\partial_{x}h - A(x,y)\partial_{y}h\right]^{2}\left[\lambda_{1}(h)\lambda_{2}(h) - Q_{+}^{2}Q_{-}^{2}\right].
\end{equation}
The metric is thus degenerate when either of the two factors vanish. Vanishing of the second factor provides the analogy of the outer horizon, although unlike for the case of point-sources here we do not have a simple guiding principle for choosing the function $h$. A natural guess would be $h=a\cosh{\eta}$ where $x=a\cosh{\eta}\cos{\psi}$ and $y=a\sinh{\eta}\sin{\psi}$.
The roots to the first factor of (\ref{linedet}) provides new degeneracies of the metric, and these are generically one-dimensional. In the case of multi-centered black holes similar, although point-like, degeneracies were found~\cite{Sundborg:multiadsbh}. They have the potentially disastrous property of not following geodesics~\footnote{These degeneracies are naked conical singularities which also have the interpretation of gravitating particles.}, but it was shown that they could be gauge transformed to follow geodesics. This would not be possible within the usual formulation in terms of a metric and the gauge invariance being diffeomorphisms. It is, therefore, a manifestation of the difference of the global structure of the diffeomorphism group of $\M$ and the gauge group in the Chern-Simons formulation~\cite{Matschull}.
The nature of the additional degeneracy of the line-source is not obvious, but it is certainly feasible that it is gauge equivalent to a minimal surface, or even to satisfy the equations of motion of a relativistic string moving in \ads.~\footnote{This is a consistency condition since otherwise this object, having tension, is unstable.}

\section{Sectors of solutions}
\label{sec:sectors}

As remarked in the introduction one needs a better understanding of the space of 
solutions with the aim of understanding the microscopic origin of the degrees of 
freedom that will saturate the Bekenstein bound of the entropy. Let us, therefore, 
more generally discuss the solutions of eqs.(\ref{bulkconstraints}) and (\ref{JtoA}).

Let us study solutions using Wilson loops. We will first consider a subclass of Wilson loops where $A=A^0T_0$ and study the gauge invariance under the $U(1)$ subgroup of
$SL(2,\RR)$ corresponding to the generator $T_0$. This case was discussed in 
\cite{Maldacena-Maoz}. We assume that we have a circular boundary at some radius $R$ which is parameterized by $x^2$. Under a gauge transformation with an element $g=\exp(-sx^2T_0)$, where $s/2\in\cal \mathbb{Z}$ by the requirement that $g$ is periodic with period $2\pi$, we have a transformation 
\begin{equation}
A_{~2}^0\longrightarrow A_{~2}^0-s.
\end{equation}
This transformation will not change the Wilson loop. However, since the 
transformation is not zero at the boundary, it is not a 
proper gauge transformation i.e. it is not compatible with the constraints (\ref{1stclconstr}). In order for this to be the case we need also to shift the zero mode of the current $J^0_0$ in accordance with (\ref{JtoA})
\begin{equation}
\label{u-one}
J_0^0(x^2)\longrightarrow J_0^0-{sk\over 2}.
\end{equation}
Requiring these values to be consistent with (\ref{unitaryrepr}) within any 
particular representation (i.e. for a fixed value $a$), we find that $k$ has to be an 
integer. Then $m$ and $m+k$ are gauge equivalent. 

We may see this in another way by 
explicitly evaluating the Wilson loop. Assuming that the trace of the Wilson loop is 
unchanged under a smooth deformation of the integration contour ${\cal C}$ to 
the boundary we have using the equality (\ref{JtoA}) of $A_2$ and $J$ 
\begin{equation}
\label{Wilson-abel}
Tr(W[{\cal C}])=Tr(e^{-\oint_{\cal C} A_{\ 2}^0(x^2)T_0dx^2})=Tr(e^{\oint_{\cal C} \frac{4\pi}{k}J^0(x^2)T_0dx^2})=2\cos({\frac{2\pi J^0_0}{k}}).
\end{equation}
We immediately see that the transformation (\ref{u-one}) indeed leaves the trace of the Wilson loop invariant for $s/2\in\mathbb{Z}$ and that $m$ and $m+k$ yield the same value for $Tr(W[C])$.

The trace of the Wilson loop will, however, be different for different values $m=0,1,\ldots, |k|-1$. Therefore, these values will give $|k|$ gauge inequivalent configurations. In addition, since $a$ in (\ref{unitaryrepr}) take two different values the total number of gauge inequivalent configurations are $2|k|$.

Let us now see what happens if we extend the discussion to gauge invariance under the full $SL(2,\RR)$ gauge group. We will first do this by considering solutions in a particular gauge. We will show that the sectors of solutions parameterized by different $J^0_0$ values, found for the Abelian subgroup, exist also for the full $SL(2,\RR)$ group, provided again $k\in \mathbb{Z}$. 

Let 
\begin{equation}A^{\ a}_i(x)=h^{\ a}_i(x), \hsv i=1,2
\end{equation} 
be any specific solution to (\ref{bulkconstraints}). Then we know by our discussion in section~\ref{sec:csanalysis} that locally we may find a gauge transformation such that the solution satisfies the gauge condition (\ref{gauge}). Denote by ${h'}^{\ a}_i(x)$ the solution in this gauge so that ${h'}_1^{\ a}(x)=0$. Then it is trivial to see that 
\begin{equation}\hat{A}^{\ a}_2(x^2)=-{4\pi\over k}(B^a_b(x^2){h'}^{\ b}_2(x^2)+B^{\prime a}(x^2)), \hsv \hat{A}_1^{\ 1}=0
\end{equation} 
is also a solution, where $B^a_b$ and $B^{\prime a}$ are arbitrary. In order for this to be a complete solution we need also
\begin{equation}
\hat{J}^a(x^2)=-{k\over 4\pi}\hat{A}^{~a}_2=B^a_b(x^2){h'}^{\ b}_2(x^2)+B^{\prime a}(x^2).
\end{equation}
These new solutions are not connected to the old ones via a gauge transformation. This 
follows since we constructed the new solutions after first completely fixing the gauge 
and the new solutions do not violate these gauge conditions. However, the new solutions 
may be connected to the old ones via some canonical transformations. As we may select the 
currents to span the physical space, it is sufficient to study field redefinitions of 
these. In order to avoid transformations that simply change the basis of the $\sltwo$ 
algebra, we fix the basis to consist of the $J_\pm={1\over \sqrt{2}}(iJ_1\mp J_2)$ and 
$J_0$ components and look for field redefinitions that preserves the algebra in this basis. 
Thus we are looking for new currents 
\begin{equation}
\label{redef}
\tilde{J}_\pm(x^2)=D_\pm(x^2) J_\pm+D^{\prime}_\pm(x^2),\hsv \tilde{J}_0(x^2)=D_0(x^2)J_0(x^2)+D^{\prime }_0(x^2),
\end{equation}
which still give an $\sltwo$ algebra.
Inserting this ansatz into the algebra (\ref{sl2}) and solving gives the unique solution 
\begin{equation}
D_-=(D_+)^{-1},\hsv D_0=1,\hsv D^{\prime}_\pm =0,\hsv D^{\prime }_0=\frac{ik}{4\pi}\frac{1}{D_+}\p_2D_+.
\end{equation}
Thus, in this basis, whenever $B^a_b$ and $B^{\prime a}$ can be written in a form that satisfies these relations then the new solutions (\ref{redef}) are canonically equivalent to the old ones. Does this mean that they are not new solutions? 
The answer is in general no. By a redefinition of a field one may change the global 
properties of the field, so that the solution has different global properties. Consider 
eg. the case when $x^2$ is an angular variable and the original currents are periodic w r t $x^2$. If $D_+$ is periodic, then the new solution $\tilde{J}$ is also periodic and, in general,
does not represent a new solution to the constraints. However, if $D_+$ is not periodic
then $\tilde{J}$ will not be periodic and then we will have a new solution. In the first case one may Fourier expand $D_+$ in terms of modes w r t $e^{-inx^2}$. Any such mode will then generate a transformation
\begin{equation}
\label{spectralI}
\tilde{J}_\pm(x^2)=e^{\mp inx^2} J_\pm,\hsv \tilde{J}_0(x^2)=J_0(x^2)+{nk\over 4\pi}.
\end{equation}
We recognize this as the spectral flow transformation which has been discussed previously 
in connection with $\sltwo$ string theories in \cite{Hwang:modinv},\cite{Hwang:modinv2} 
and \cite{Maldacena:wzw1}. As far as the classical discussion here we see that the 
spectral flow transforms from one solution to what appears to be a physically identical solution  (in contrast to the string case). We will see shortly that this is an almost correct conclusion.

In the case that $D_+$ is not periodic, we have different possibilities. Either we have 
that $(D_+)^N$ is periodic for some integer $N>1$ or it is not. If it is then we may 
Fourier expand $D_+$ in terms of modes w r t $e^{-inx^2/N}$. Any such mode will generate a 
transformation 
\begin{equation}
\label{spectralII}
\tilde{J}_\pm(x^2)=e^{\mp inx^2/N} J^\pm,\hsv \tilde{J}_0(x^2)=J_0(x^2)+{nk\over 4\pi N}.
\end{equation}
In this case the transformations are not the spectral transformations discussed in the 
context of $\sltwo$ string theories. However, as far as the resulting symmetry algebra the 
transformed currents represent perfectly well-defined objects. By a Fourier expansion one 
finds an algebra of the modes which may be thought of as part of a twisted $\sltwo$ algebra. Interestingly enough similar spectral transformations appear in connection with string theory on non-compact orbifolds~\cite{Dabholkar1, Dabholkar2, MartMcElg, Son, KeskHem}.

Let us now find these solutions without referring to a specific gauge. Consider the transformation
\begin{equation}
\label{gauge-spectral}
\tilde{A}=g^{-1}dg+g^{-1}Ag,\hsv g=exp(-sx^2T_0).
\end{equation}
This transformation yields explicitly
\begin{equation}
\tilde{A}^{\ 0}_\mu=A^{\ 0}_\mu-s\delta_{\mu,2},\hsv \tilde{A}^{\ \pm}_\mu=e^{\mp isx^2}A^{\ \pm}_\mu,
\end{equation}
where $A^\pm=iA^1\mp A^2$.
Let $x^2$ be an angular variable with period $2\pi$, then the transformation is a proper 
bulk gauge transformation, i.e. $g$ is periodic with period $2\pi$ if 
$s/2\in \mathbb{Z}$. The transformation is simply a rotation in the components $A^1$ and $A^2$ with angle $sx^2$ and a shift in the $A^0$ component. The trace of the Wilson loop is invariant under this 
transformation. However, it is not a proper gauge transformation on the boundary as $g\neq 0$ there. Completing this transformation with 
\begin{equation}
\label{spectral}
\tilde{J^0}=J^0+{sk\over 4\pi},\hsv \tilde{J}^\pm=e^{\mp isx^2}J^\pm,
\end{equation}
then we have that the new fields are consistent with the constraints 
(\ref{1stclconstr}). We recognize again the spectral flow transformation on the 
currents, which however are not gauge-transformations.
Since the trace of the Wilson loop is unchanged we will not change the physical content of 
the solution through the transformation. We should then require that the 
transformation (\ref{spectral}) will transform from one particular unitary 
representation to an isomorphic one i.e. the Casimir is unchanged as is the value of 
$a$ in (\ref{unitaryrepr}). This is certainly the case for the BTZ black hole as can 
be seen from (\ref{BTZWilson}) since the trace of the Wilson loop only depends on the 
Casimirs. Comparing (\ref{unitaryrepr}) and (\ref{spectral}) we 
see that equivalent solutions will belong to the isomorphic representation 
provided $k$ is an integer.  On the other hand if $s/2$ is not an integer, then the 
spectral flow will, in general, change the Wilson loop, and this is a new solution.
Requiring these new solutions to be consistent with (\ref{unitaryrepr}) 
one finds that there are exactly $2|k|$ new sectors of 
solutions corresponding to the values $s=1/|k|, 2/|k|,\ldots, 2-1/|k|$. This is the sought 
after generalization of the sectors of solutions discussed for the simple $U(1)$ subgroup.

For example, the Wilson loop enclosing the origin at fixed time may be written 
\begin{equation}
\label{WilsonU}
Tr(W)=Tr\left(U^{-1}(x^0,x^1,x^2=0)U(x^0,x^1,x^2=2\pi)\right),
\end{equation}
where $U$ is an element of the gauge group such that $A=U^{-1}dU$ (which always exists 
locally). Under the transformation (\ref{gauge-spectral}) we have
\bea
Tr(W)\longrightarrow Tr(\tilde{W})=Tr\left((Ug)^{-1}(x^0,x^1,x^2=0)(Ug)(x^0,x^1,x^2=2\pi)
\right).
\eea
Hence, if $g(x^0,x^1,x^2=2\pi)g^{-1}(x^0,x^1,x^2=0)\neq 1$ then, in general, 
$Tr(W)\neq Tr(\tilde{W})$.

As an explicit example consider the BTZ solution. The Wilson loop 
centered around the origin at $x^1=R>r_+$ was calculated in (\ref{BTZWilson}). Under the 
transformation (\ref{gauge-spectral}) this changes as
\begin{equation}
Tr(e^{-\oint \tilde{A}})=Tr(e^{(1+{\cal J}){(2\pi r_+/ l)}T_2}e^{-2\pi sT_0}).
\end{equation}
This may be evaluated to give
\begin{equation}
\label{newBTZWilson}
Tr(e^{-\oint \tilde{A}})=2\cos(\pi s)\cosh\left[{\pi \over l}(r_+-r_-)\right].
\end{equation}
For $s/2\in \mathbb{Z}$ we recover the original value, but for other values the trace of 
the Wilson loop has changed. We see again by using (\ref{btzreps}) that the trace of the 
Wilson loop only depends on the Casimir of the representation and so the new sectors of 
solutions all belong to isomorphic representations. Comparison with (\ref{Wilson-abel}) 
shows that the our generalization for the non-Abelian case here is a natural extension of 
the Abelian one. 

We can rewrite (\ref{newBTZWilson}) in the form
\begin{equation}
\label{newBTZWilson1}
Tr(e^{-\oint \tilde{A}})=\cosh\left[{\pi }({r_+-r_-\over l}+is)\right]+\cosh\left[{\pi }
({r_+-r_-\over l}-is)\right].
\end{equation}
Configurations for which the charge $q=(r_+\mp r_-)/l$ is purely imaginary corresponds to 
naked singularities \cite{Banados-H-T-Z}. The new solution resembles, therefore, a sum 
of two configurations, each of which is a mixture of a BTZ black hole solution and a naked 
singularity of charge $q=\pm s$.

In fact, due to this simple formula for the Wilson loop, it is readily realized that one 
may produce any given solution with static sources by a non-proper gauge transformation. Starting from the CS vacuum 
solution $A=A'=0$ we perform a time-independent gauge transformation
\begin{equation}
   \label{solgen}
   A\rightarrow \tilde{A} = h^{-1}dh
\end{equation}
where $h=e^{f(x^{1},x^{2})T_{2}}$ for some function $f(x^{1},x^{2})$. Then
\begin{equation}
   \label{Wilsonsolgen}
   Tr\left(W[C]\right) = Tr\left(h^{-1}(x^1,x^2)h(x^1,x^2+C)\right) \equiv 2\cosh{\mathcal{F}[C]}
\end{equation}
for a closed loop $C$.
By choosing $h$ such that
\begin{equation}
   \label{curvefn}
   \frac{\del \mathcal{F}[C]}{\del C} = 0
\end{equation}
everywhere, except for $\del$-singularities when $C$ moves over a source (point or line), we can construct 
any charge distribution. Since, in the bulk, the solutions are completely classified by 
the Wilson loops, and hence the CS charge distributions, we can reconstruct any given 
solution by a suitable choice of $h$. For example, by choosing $h=e^{x^{2}T_{2}(r_{+}-r_{-})/l}$ we have $\mathcal{F}[C] = \pi(r_{+}-r_{-})/l$, except at the origin where it is ill defined. Hence (\ref{curvefn}) is satisfied except at the origin where we have a point source. This is the BTZ black hole (cf. (\ref{BTZWilson})).

Let us now compare with our previous discussion using a particular gauge. Comparison 
between (\ref{spectralII}) and (\ref{spectral}) shows that they coincide provided $n$ 
in the former equation is {\em an even integer} i.e. classically equivalent solutions 
are identified through the spectral flow if they have the same periodicity under 
rotations $4\pi$. If we have a solution that has a certain periodicity, then there 
will, in general,  exist an inequivalent solution with the same periodicity. The 
solutions are connected through the transformation (\ref{spectral}) with $s=1$. We 
see from (\ref{spectral}) that the transformation changes the values of $J^0_0$ from integer (half-integer) valued representation to half-integer (integer), i.e. the value of $a$ in (\ref{unitaryrepr}) is changed.

Under the transformation (\ref{spectral}) the Hamiltonian for the 
$\sltwo$ model changes by (we display only one chirality)
\bea
H={2\pi\over kl}\int dx^2(2J_+(x^2)J_-(x^2)+(J_0)^2)=\nonumber\\
+{2\pi\over kl}\int dx^2(2\tilde{J}_+(x^2)\tilde{J}_-(x^2)+(\tilde{J}_0(x^2)+{sk\over 
4\pi})^2),
\eea
which shows that for $\tilde{J}_0^0=a+m-{sk/ 2}$ the Hamiltonian is form invariant and 
unchanged under the transformation for general values of $s$, not only integers. 
Furthermore, the Hamiltonian remains local.

Let us also discuss the role of the boundary in the presence of several sources. If all 
sources are confined to a finite region of space, then one may choose a boundary that 
encircles all the sources. However, we may also choose the boundary such 
that it consists of several disjoint pieces, each piece encircling a source. Our 
manifold would consist of all space outside these boundaries. The formalism in
section~\ref{sec:csanalysis} then gives that we should introduce a current $J^a$ on each boundary, that depends on the local parameter that describes the boundary. Clearly, since each current has in general a different space-time dependence, they are not the same currents. However, they are not independent of each other. To see how they are connected we impose the gauge (\ref{gauge}). This implies the solution (\ref{solveingauge}), which seems like a contradiction since the identification is true near every boundary and $x^2$ can, in general, not be chosen to parameterize every boundary. The resolution is that the gauge (\ref{gauge}) is only, in general, possible to enforce locally or, equivalently, if we enforce it globally, then $A_2$ cannot exist as a smooth function globally. 

Picture now a solution with several sources, each source belonging to a different sector. Encircle every source by a boundary. Then on these boundaries we have currents, each of them having a different locality property. In this case it is clear that our degrees of freedom consist of several independent currents. If we assume that we can smoothly deform the boundaries and join them into a single boundary, this boundary is associated with all these currents superimposed. {\em At the quantum level this corresponds to constructing the states as tensor products of states belonging to different sectors i.e. the second quantized~\footnote{We use the expression second quantized here in analogy with second quantization of string theory. It seems that to describe all possible configurations we need to construct the "string field theory" corresponding to the boundary CFT.} theory is constructed by tensoring the different sectors.}

Let us now discuss the most important aspect, namely the locality properties. Firstly, 
even if the dreibein is non-local, the metric does not have to be.
If we choose to transform $A'$ with the same element $h$ as $A$, the resulting metric will 
have local boundary conditions even when $A$ and $A'$ do not. Thus we see that the 
locality properties are invariant under diagonal gauge transformations.
But we claim that one must also include solutions where $A$ and $A'$ belong to different 
sectors. Normally one would not consider such solutions, since they do not have the 
desired locality properties associated with space-time. {\em Our proposal is that it is 
only the resulting final state in the second quantized theory that needs to 
be local, not necessarily each of the states in the first-quantized version.} This means 
specifically that we form the states in the second-quantized theory by first tensoring 
states of each sector and then applying a projection to local states. In classical terms 
these solutions are associated with several sources where the non-localities of each 
source cancel out, yielding a 
local solution. In particular, this means that the only one-source solution that we 
include in the spectrum is the local BTZ solution.

A slightly different way of viewing the solutions described above is via the relation to a twisted \sltwo already mentioned.
For currents $J^{a}(x^{2})$ with non-local periodicities, i.e. $J^{a}(x^{2} + 2\pi N) = J^{a}(x^{2})$ where $N\neq 1$, a decomposition into Fourier modes is performed in the basis $\{e^{inx^{2}/N}\}_{n\in\mathbb{Z}}$. Labelling the corresponding modes by \[ J^{a}_{n/N} = J^{a}_{m + s/N} \] with $m\in\mathbb{Z}$ and $s\in\mathbb{Z}\ \mbox{mod}\ N$, with the understanding that in our case $N=|k|$, these satisfy the (quantized) relations
\begin{eqnarray}
   \left[ J^{+}_{m+s/N},J^{-}_{n+s'/N} \right] & = & J^{0}_{m+n+(s+s')/N} + (m+s/N)\frac{k}{2}\delta_{m+n+(s+s')/N} \nn \\
   \label{twistedsl2}
   \left[ J^{0}_{m+s/N},J^{\pm}_{n+s'/N} \right] & = & \pm J^{\pm}_{m+n+(s+s')/N} \\
   \left[ J^{0}_{m+s/N},J^{0}_{n+s'/N} \right] & = & (m+s/N)\frac{k}{2}\delta_{m+n+(s+s')/N}\nn
\end{eqnarray}
i.e. a twisted \sltwo algebra of order $N$.
The redefinition \[ K^{a}_{n} = J^{a}_{n/N} \] displays the isomorphism with an ordinary \sltwo algebra of level $\tilde{k}=k/N$
\begin{eqnarray}
   \left[ K^{+}_{m},K^{-}_{n} \right] & = & K^{0}_{m+n} + m\frac{\tilde{k}}{2}\delta_{m+n} \nn \\
   \label{tildesl2}
   \left[ K^{0}_{m},K^{\pm}_{n} \right] & = & \pm K^{\pm}_{m+n} \\
   \left[ K^{0}_{m},K^{0}_{n} \right] & = & m\frac{\tilde{k}}{2}\delta_{m+n}.\nn
\end{eqnarray}
A corresponding Virasoro algebra of central charge
\begin{equation}
   \label{twistedc}
   \tilde{c}=\frac{3\tilde{k}}{\tilde{k}+2} = \frac{3k}{k+2N}
\end{equation}
is constructed in the usual way via the Sugawara bilinear
\begin{eqnarray}
   L_{n} & = & \frac{1}{\tilde{k}+2}\sum_{m\in\mathbb{Z}}:\left[ K^{+}_{m}K^{-}_{n-m}+K^{-}_{m}K^{+}_{n-m}+K^{0}_{m}K^{0}_{n-m}\right]:\nn\\
      & = & \frac{1}{\tilde{k}+2}\sum_{m\leq -1}\left[ K^{+}_{m}K^{-}_{n-m}+K^{-}_{m}K^{+}_{n-m}+K^{0}_{m}K^{0}_{n-m}\right]\nn\\
      \label{twistedsugawara}
      & + & \frac{1}{\tilde{k}+2}\sum_{m\geq 0}\left[ K^{-}_{n-m}K^{+}_{m}+K^{+}_{n-m}K^{-}_{m}+K^{0}_{n-m}K^{0}_{m}\right].
\end{eqnarray}

The twisted algebra (\ref{twistedsl2}) contains $N$ sub \sltwo algebras of level $k$ generated by
\begin{eqnarray}
   J^{(s)\pm}_{m} & = & J^{\pm}_{m\mp s/N} \nn \\
   \label{subsl2}
   J^{(s)0}_{m} & = & J^{0}_{m}-\frac{s}{N}\frac{k}{2}\delta_{m}
\end{eqnarray}
for $s=\{0,1,\ldots,N\}$, and each of these yields a Virasoro algebra of central charge $c=\frac{3k}{k+2}$ via the Sugawara construction
\begin{eqnarray}
   \label{subsugawara}
   L^{(s)}_{n} & = & \frac{1}{k+2}\sum_{m\in\mathbb{Z}}:\left[J^{(s)+}_{m}J^{(s)-}_{n-m} + J^{(s)-}_{m}J^{(s)+}_{n-m} + J^{(s)0}_{m}J^{(s)0}_{n-m}\right]:
\end{eqnarray}
The generators $J^{(s)a}_{m}$ are obtained by a generalized spectral flow transformation starting from an untwisted \sltwo algebra, the generalization consisting of allowing non-integral $s$. The expressions (\ref{subsl2}) are also recognized as the mode-versions of the relations (\ref{spectralI}), where they were considered as solution-generating transformations.

\section{The black hole entropy}
\label{sec:entropy}

Using the analysis of the preceding sections we now proceed to consider the quantum 
theory and, in particular, the entropy calculation. We will restrict ourselves to the case when the boundary is a circle parameterized by $x^2$ with period $2\pi$. Our degrees of freedom are the two \sltwo currents, which by a Fourier decomposition, defines a standard state space of the 
$\sltwo$ conformal field theory. The ground states are defined to be states 
$|R\rangle|R'\rangle$ which transform in some unitary representation of 
$SL(2,\RR)\times SL(2,\RR)$ and that satisfy
\begin{equation}
\label{g-state}
J^a_{p}|R\rangle|R'\rangle= J^{\prime a}_p|R\rangle|R'\rangle=0,\hsv p>0.
\end{equation}
The principal continuous 
representations ${\cal C}^a_j$ of $SL(2,\RR)$ have a quadratic Casimir  
$C^{(2)}=-j(j+1)$, $j=\half+i\rho$, $\rho\in\RR$, and by comparing with (\ref{btzreps}) 
and (\ref{btzreps3}) we see that the BTZ solution belongs to this class of representations. The same is true for all the examples of solutions (except adS) given in 
the previous sections.
Building on the discussion of the new sectors of solutions we take the state-space to be a tensor product of $2|k|$ parts, each built from a ground state $|\rho\rangle_{s}$ such that
\begin{equation}
\label{g-state-s}
J^{(s)}_{p}|\rho\rangle _s|\rho'\rangle _s= J^{\prime (s)}_p|\rho\rangle _s|\rho'\rangle _s=0,\hsv p>0.
\end{equation}
Here the index $s=0,1,\ldots ,2|k|-1$ labels the different sectors of (\ref{spectralI}) and $\rho$ and $\rho'$ are the Casimir eigenvalues (\ref{btzreps}) and (\ref{btzreps3}).
The ground state of the full theory is then
\begin{equation}
   \label{tensorgroundstate}
   \bigotimes_{s=0}^{2|k|-1}|\rho\rangle_{s}|\rho'\rangle_{s}.
\end{equation}

The Hamiltonian for the full theory is a sum over all sectors
\bea
   H & = & \sum_{s=0}^{2|k|-1}H^{(s)} \nonumber\\
   & = & \sum_{s=0}^{2|k|-1}{2\pi\over (k+2)l}\int dx^2:(J^{(s)+}(x^2)J^{(s)-}(x^2)+J^{(s)-}(x^2)J^{(s)+}(x^2)+(J^0)^2):\nonumber\\ 
   & + &\ldots(\mbox{the other chirality}).
\eea
Defining $L_0$ as $l$ times the terms associated with one chirality and $\bar{L}_0$ analogous for the other chirality we have
\begin{equation}
L_0={1\over k+2}\sum_{s=0}^{2|k|-1}\sum_{n\in\mathbb{Z}}:\left[J_{-n}^{(s)+}J_{n}^{(s)-}+J_{-n}^{(s)-}J_{n}^{(s)+}+J_{-n}^{(s)0}J_{n}^{(s)0}\right]:.
\end{equation}
If we assume each sector, described by $J_{n}^{(s)a}$, to be independent, i.e. the current modes of different sectors commute, the total conformal anomaly $c$ is the sum of $2|k|$ terms each having the value $c^{(s)}=3k/(k+2)$ so that $c=-6k^2/(k+2)$. For large $|k|$ this reduces to $c=6|k|$, which by (\ref{level}) coincides with the conformal charge found by Brown and Henneaux \cite{Brown-Henneaux}.

As we discussed in the preceding section, the physical states are defined to be those which are local. To define local states we introduce a "locality index" $\mathcal{I}$, which for excited states in the sector $s$ has value $s/|k| \mod |k|$. For states which have excitations in several sectors $\mathcal{I}$ is defined to be the sum of the indices for each sector. Thus for a state which has excitations in sector $p$ and $q$, we have $\mathcal{I}=\frac{p+q}{|k|} \mod |k|$. A local state is defined to be a state with $\mathcal{I}=0$, and the physical states are defined as the projection down to the $\mathcal{I}=0$ states. The number of states in the local subspace is a factor $\frac{1}{2|k|}$ less than in the full state space.

The calculation of the entropy is straightforward, as each sector has a state space that is irreducible and the number of excited states is the same as those of three free field state spaces. Thus, for fixed values of $\rho$ and $\rho'$, and hence of the Casimirs, the asymptotic number of states with (large) eigenvalues $\Delta$ and $\bar{\Delta}$ of $L_0$ and $\bar{L}_0$ is
\begin{equation}
   \label{density}
   \varrho(\Delta,\bar{\Delta})\sim e^{2\pi\sqrt{6|k|\Delta\over 6}}e^{2\pi\sqrt{6|k|\bar{\Delta}\over 6}}.
\end{equation}
In order to make contact with the entropy of the black hole one uses the relations (\ref{btzreps}) and (\ref{btzreps3}) for the BTZ case, which imply
\begin{equation}
\label{Lvalue}
\Delta={1\over |k|}\left({k(r_+-r_-)\over 2l}\right)^2,\hsv \bar{\Delta}={1\over |k|}
\left({k(r_++r_-)\over 2l}\right)^2
\end{equation}
and assumes that they also apply, at least for large $|k|$, in the general case. Inserting this into (\ref{density}) and using (\ref{level}) we find the Bekenstein-Hawking expression for the entropy
\begin{equation}
S=\ln\varrho(\Delta,\bar{\Delta})={2\pi r_+\over 4G}.
\end{equation}
The effect of the locality projection is merely an additive logarithmic correction $\sim \log{|k|}$.

As mentioned in the previous section there also exists a slightly different possibility, namely to choose our sectors not to be independent, but related as parts of a product of two twisted representations of order $|k|$.
The calculation of the entropy is completely analogous and yields, to leading order, the same result.
It is not clear to us which is the correct choice. To determine this probably requires a more careful analysis of the classical solutions.

Let us end this section by discussing another derivation of the entropy, namely Carlip's first calculation~\cite{Carlip95}. This derivation seems at first sight very different from all the others. However, we will see in our formulation that the differences boil down to three principal points. 
The main difference is the choice of ground state.
The ground state is chosen such that in one chiral sector the creation operators are switched with the annihilation operators, as compared to the conventional choice. This has the implication that the role of the level $k$ is switched with $-k$ in one of the sectors.
The signs of the Casimirs still imply that the principal continuous representations are to be used, while $\rho$ and $\rho'$ are related to the black hole parameters as
\begin{equation}
   \label{radiirep}
   r_{\pm} = \frac{l}{2|k|}(\rho'\pm\rho) \equiv \frac{l}{2|k|}\rho_{\pm}.
\end{equation}
With this choice of ground state the mode number operator is no longer related to $L_{0}+\bar{L}_{0}$, but rather to $L_{0}-\bar{L}_{0}$ which has an expectation value of the form
\bea
   \langle L_{0}-\bar{L}_{0}\rangle & = & -\frac{\rho^{2}-\frac{1}{4}}{k+2} + \frac{\rho^{'2}-\frac{1}{4}}{k-2} + N + N'\nonumber\\
   \label{2chHvev}
   & = & \frac{1}{k^{2}-4}\left[\left(\rho_{-}+\frac{k}{2}\rho_{+}\right)^{2}-1\right] - \frac{\rho_{+}^{2}}{4} + N + N'
\eea
where $N$ and $N'$ are the mode numbers in the respective sectors.
Classically, the quantity $L_{0}-\bar{L}_{0}$ corresponds to the doubly chiral Hamiltonian (\ref{2chiralH}), which has the following physical meaning
\begin{equation}
  \label{2chHvalue}
   H_{\tt chir-chir} = \frac{J}{l}.
\end{equation}
The other differences in this approach are that we now keep $L_{0}-\bar{L}_{0} = J$ and $\rho_{+}$ fixed in (\ref{2chHvev}). In the limit of large $|k|$ one can evaluate the value of $N_{tot}=N+N'$ which by (\ref{2chHvev}) is given by 
\begin{equation}
   N_{tot} = J - \frac{1}{k^{2}-4}\left[\left(\rho_{-}+\frac{k}{2}\rho_{+}\right)^{2}-1\right] + \frac{\rho_{+}^{2}}{4}.
\end{equation}
Inserting this into $\varrho(N_{tot}) = e^{\pi\sqrt{N_{tot}}}$, which is the number of states in this case, and integrating over $\rho_{-}$, one finds for small $J$ the Bekenstein-Hawking entropy.
Thus, in this approach the radius of the outer horizon $r_{+}$ is held fixed, and also the expectation value of $L_{0}-\bar{L}_{0}$.
In the original paper~\cite{Carlip95}, the integration over $\rho_{-}$ is an integration over the zero mode of the spin connection on the horizon. Here it instead corresponds to integrating over a continuous representation parameter, a very natural thing to do.
An unnatural feature, however, is that in keeping $L_{0}-\bar{L}_{0} = J$ fixed but letting $\rho_{-}$ vary, we classically allow the mass $M$ to vary. The calculated entropy does, therefore, {\em not} correspond to a black hole of a certain mass.
The main reason, however, why we believe that this approach is not the correct one, is that due to the choice of ground state the theory is effectively a chiral one. As we have seen, this is not the choice implied by gravity.

\section{Discussion}
\label{sec:discussion}

Let us summarize the main results.
The first one is the gauge invariant relation between gravitational solutions, in the presence of a boundary, and solutions of the $SL(2,\mathbb{R})$ WZNW model. The property of gauge invariance is something we would like to emphasize. In particular, the important consequence that the correspondence is independent of the particular boundary condition applied to the gauge fields.
A second important insight is that the state space of the \sltwo algebra necessarily contains negative norm states. This is a disappointing feature, and one which could be used to argue against the relevance of our results. Nevertheless, this is inevitable following our, quite rigorous, line of arguments.
The final, and also the central result is our proposal that in the second quantized theory we should include $2|k|$ individually non-local solutions and apply a projection to local states. This was shown to be just enough to account for the Bekenstein-Hawking entropy, and we conjecture that this is enough to describe all possible solutions.

There are also a number of open questions.
For instance, explicitly constructing the dictionary between solutions of the CS theory and solutions of the WZNW model. This calls for a better geometric understanding of the multitude of solutions, both in each sector individually and for solutions involving several sectors. It would be interesting to understand from a geometric point of view, why there are exactly $2|k|$ sectors. Intuitively, the answer one would expect is that if there are more than $2|k|$ sources, the various phases add up to an integer phase plus a phase between $0$ and $|k|$. This may also answer the question why we need only unitary representations of $SL(2,\mathbb{R})$ and not of any cover of this group. 
A better understanding of the classical solutions may also be necessary to determine whether taking the different sectors to be independent, or to form two twisted \sltwo algebras is the correct choice.

The matter of unitarity is another question which needs further investigation. Without any modification, the theory is not unitary. One might consider constraining the theory further by applying extra conditions. Modding out the two timelike $U(1)$ subgroups would for instance help. The trick is to find constraints which make sense in the gravitational interpretation. A serious downside to this approach is that the entropy would no longer be the Bekenstein-Hawking value. The number of timelike modes is simply too large.

Our physical state space was constructed by a projection onto local states. Even in this projection, however, there may still be many physically equivalent states. Thus it is conceivable that further projections are needed, and in some respects this even looks natural. From the naive picture of the sectors of states presented in section~\ref{sec:sectors} it seems obvious that permutations of the factors in the state space transform between physically equivalent states. In this respect it appears natural to project onto the symmetrized product. Another point is that there is an apparent mismatch between inequivalent classical gravitational solutions, and inequivalent solutions in the full theory described by $A$, $A'$ and $J$. Since the metric is bilinear in the dreibein $e^{a} = \frac{1}{2}(A^{a}-A^{'\, a})$ it is invariant under diagonal gauge transformations, i.e. under gauge transformations belonging to the vector subgroup $SL(2,\mathbb{R})_{V}$, of $SO(2,2)$. By utilizing a non-local diagonal transformation of the form (\ref{solgen}), however, we find solutions corresponding to identical metrics but which do not yield the same physical state according to the arguments in section~\ref{sec:sectors}.
This implies that one should project even further.
A better understanding of the classical solutions may answer these questions.

The construction of the Hilbert space presented here shares some similarities with certain constructions in string theory. Firstly the one pointed out in section~\ref{sec:sectors} to string theory on non-compact orbifolds~\cite{Dabholkar1, Dabholkar2, MartMcElg, Son, KeskHem}, and furthermore the approach to second quantized string theory which goes under the name matrix strings. In this case the second quantized Hilbert space is a tensor product of single-string Hilbert-spaces, and an action of the symmetric group is factorized out. The resulting theory is called a symmetric product CFT, and it is an example of a permutation orbifold CFT.
Interestingly enough the string theory description of the \dd system which describes the BTZ black hole in a string theory setting is believed to be equivalent via the AdS/CFT correspondence to a symmetric product CFT. Since string theory has a large matter content it should not be equivalent to the model described here, which contains only the gravitons. One of the most interesting continuations of this project would be to investigate whether the gravitational state space is embedded in the state space of the \dd system. This is an important question from the point of unitarity, since the \dd system is believed to be described by a unitary CFT.

It is not obvious how our results relate to the AdS/CFT correspondence. One feature which seem to differ is the appearance of the conformal group on the boundary. In the AdS/CFT correspondence the conformal group on the boundary arises as the isometry group of the bulk theory. Here, however, the theory on the boundary has a local $SL(2,\mathbb{R})$ invariance which contains more than just conformal transformations. Instead, conformal transformations are generated by bilinears in the generators of the local $SL(2,\mathbb{R})$ via the Sugawara construction.
We suspect that this difference may date from the way these two correspondences are formulated. The AdS/CFT relation is formulated much like a correspondence between two field theories on a fixed background. On the contrary, the relation described here does not presume a certain background. There is no direct correspondence between the three-manifold on which the CS theory is defined and the spacetime described by the metric, and, as a consequence, we allow arbitrary fluctuations of the metric. In the conventional approach to the AdS/CFT correspondence on the other hand, one considers linearized gravity in an AdS background. In a sense the AdS/CFT formulation looks like a gauge fixed version of the CS/WZNW formulation.

There are also some less acute questions, but which are nevertheless interesting.
One is the status of the line-source solutions. Line sources have been considered previously in a flat background~\cite{DesJack}. Although they are one-dimensional, they are not the conventional three-dimensional black string solutions~\cite{blackstring1}. The known three-dimensional black-string solutions of supergravity are infinitely extended, while the specific solution presented here is open and of finite (arbitrary) length. The first black string solution, found by Horne and Horowitz, is furthermore asymptotically flat, while our solution is asymptotically BTZ. There are BPS solutions found in $\mathcal{N}=(2,0)$ supergravity~\cite{blackstring2}, however, which are asymptotically \ads.

One may also speculate about two-dimensional sources, i.e. membrane solutions. Nothing apparent prevents the construction of such solutions. The fact that gravitons do not propagate in the bulk seems to imply, however, that these solutions will not provide new input in the present context. 
For all extended solutions one should also investigate stability.

Recently there has been a surge in interest of gravity on de-Sitter (dS) spacetime, and a correspondence between gravity, or string theory, in a dS background and a CFT on one of the conformal (spacelike) boundaries have been proposed~\cite{Strominger:dSCFT}. It is possible to rewrite the Einstein-Hilbert action also with a positive cosmological constant in terms of a CS theory~\cite{Witten:gravcs}, but the formulation used in this paper does not seem suitable to investigate this relation. The reason is that the conformal boundary of dS is spacelike, and this makes a Hamiltonian treatment more obscure. One could still imagine, however, using this formulation to investigate the space of solutions in a manner analogous to the work presented here. We hope to return to this question in the future.

The construction of the \sltwo state space suggests another possible avenue. This construction is, as already mentioned, quite different from the representation space conjectured to be relevant for string theory on \ads~\cite{Hwang:modinv, Maldacena:wzw1}. In the string case it seems one must include all (integer) spectrally flowed representations, while here we make use of only $2|k|$ (non-integer) spectral flows. Maybe this provides a second way to describe also string theory on \ads. For this to be an option, it must be possible to find modular invariant partition functions using only characters corresponding to the $2|k|$ twisted representations. Such a partition function would have one great advantage, namely that there would be no divergences corresponding to the sum over all spectral flows and zero-mode eigenvalues. One objection to this possibility is that it seems difficult to consider this string theory as part of, say, a compactification from ten dimensions. In such a setting one apparently needs all spectral flows for consistency due to the upper bound of the $L_{0}$ eigenvalue in each spectral flow sector.

\vs{10mm}

\noi
{\bf Acknowlegement:} We would like to thank Teresia M{\aa}nsson and Bo Sundborg for interesting discussions. S.H. is in part supported by the Swedish Research Council.

\vs{2cm}
\section{Appendix: Conventions}

$SL(2,\RR)$-algebra
\begin{equation}
[T_a,T_b]=f^c_{~ab}T_c,\ f_{012}=\epsilon_{012}=1.
\end{equation}
\begin{equation}
Tr(T_a)=0,\hsv Tr(T_aT_b)=\frac{1}{2}\eta_{ab},\hsv Tr(T_0T_1T_2)={1\over 4},  \hsv \eta_{ab}={\tt diag}(-1,+1,+1)
\end{equation}
\begin{equation}
(T_a)^\ast=T_a,\hsv  \eta_{ab}={\tt Killing metric}= \frac{1}{2}f^c_{ad}f_{bc}^d
\end{equation}
Explicit basis
\begin{equation}
T_0=\half\left(\begin{array}{rr}0&-1\\1&0\end{array}\right),\hsv T_1={1\over 2}\left(\begin{array}{rr}1&0\\0&-1\end{array}\right),\hsv T_2={1\over 2}\left(\begin{array}{rr}0&1\\1&0\end{array}\right)
\end{equation}
\begin{equation}
T_1T_2=-\half T_0,\hsv T_2T_0=\half T_1,\hsv T_0T_1=\half T_2.
\end{equation}
Chern-Simons action
\begin{equation}
   I_{CS}[A] = - \frac{k}{4\pi}\int_{\M}Tr\left[A\wedge dA + \frac{2}{3} 
   A\wedge A\wedge
   A\right]
\end{equation}
Basic Poissonbrackets
\begin{equation}
\{A_{2a}(x),A_{1b}(x')\}=-\frac{4\pi}{k}\eta_{ab}\delta^2(x-x')
\end{equation}
\begin{equation}
\{J_a(\phi),J_b(\phi')\}=-f_{~ab}^{c}J_c(\phi)\delta(\phi-\phi')-
\frac{k}{4\pi}\eta_{ab}\partial_\phi\delta(\phi-\phi')
\end{equation}
\begin{equation}
[J_a(\phi)]^\ast=J_a(\phi)
\end{equation}
Constraint generators
\begin{equation}
G'_\la=\int_D d^2x(\p_1A_{2a}-\p_2A_{1a}+f_{abc}A^b_1A^c_2)\la^a-\int_{\p D} 
d\phi(A_{2a}+\frac{4\pi}{k}J_a)\la^a
\end{equation}
\begin{equation}
\{G'_\la,G'_{\la^\prime}\}=G'_{\la^{\prime\prime }},\hspace{10mm} 
\la^{\prime\prime c}=\frac{4\pi}{k}\ep_{ab}^{~~c}\la^a\la^{\prime b}
\end{equation}
Fourier decomposition. For $\phi\in[0,2\pi N]$ i.e. period of $2\pi N$
\begin{equation}
\delta(\phi)=\frac{1}{2\pi N}\sum_{m\in \cal{Z}}e^{-i\frac{m}{N}\phi}\hspace{10mm}
\int_0^{2\pi N}d\phi\delta(\phi)=1
\end{equation}
\begin{equation}
J_a(\phi)=\frac{1}{2\pi N}\sum_{m\in 
\cal{Z}}J_{am}e^{-i\frac{m}{N}\phi}\hspace{10mm}J_{am}=\int_0^{2\pi N}d\phi 
J_a(\phi)e^{i\frac{m}{N}\phi}
\end{equation}
\begin{equation}
[J_a(\phi)]^\ast=J_a(\phi), \hs{10mm} (J_{am})^\ast=J_{a-m}
\end{equation}
Algebra in modes
\begin{equation}
\{J_{am},J_{bn}\}=-f_{~ab}^{c}J_{c\hspace{.2mm}m+n}+i\frac{k}{2}m\eta_{ab}
\delta_{m+n,0}
\end{equation}
Connection between $A_{2a}$ and $J_a$: If we have a field configuration that satisfies 
$
\p_1A_{2a}-\p_2A_{1a}+\ep_{abc}A^b_1A^c_2=0
$
and $A_{2a}(\phi)=A_{20}+\mbox{ (non-constant parts)}$ at the boundary, then at the boundary
\begin{equation}
A_{20}=-\frac{2}{k}J_{a0}
\end{equation}
Energy-momentum tensor
\begin{equation}
T(\phi)=-\frac{2\pi N}{k}\eta^{ab}J_a(\phi)J_b(\phi)=\frac{1}{2\pi N}\sum_{m\in 
\cal{Z}}T_{m}e^{-i\frac{m}{N}\phi}
\end{equation}
\begin{equation}
T_m=\int_0^{2\pi N}d\phi 
T(\phi)e^{i\frac{m}{N}\phi}=-\frac{1}{k}\sum_{m\in 
\cal{Z}}\eta^{ab}J_{a\hs{.2mm}m-n}J_{bn}
\end{equation}
Algebra
\begin{equation}
\{T_\la,T_\mu\}=T_{\la\p\mu}-T_{\mu\p\la}, \hs{10mm}T_\la=\int_0^{2\pi N}d\phi 
T(\phi)\la(\phi)
\end{equation}
\begin{equation}
\{T_m,T_n\}=-i(m-n)T_{m+n}
\end{equation}
Hamiltonian
\begin{equation}
H=\int_0^{2\pi N} T(\phi)d\phi =T_0=-\frac{2\pi N}{k}\int_0^{2\pi N} \eta^{ab}J_a(\phi)J_b(\phi)d\phi=
-\frac{1}{k}\sum_{m\in 
\cal{Z}}\eta^{ab}J_{a\hs{.2mm}-n}J_{bn}
\end{equation}
Equations of motion:
\begin{equation}
(\partial_0+\partial_\phi)J_a(t,\phi)=0\ \Rightarrow\ J_a(t,\phi)=J_a(t-\phi)\ \Rightarrow\ J_a(\phi)=\frac{1}{2\pi N}\sum_{m\in 
\cal{Z}}J_{am}e^{i\frac{m}{N}(t-\phi)}
\end{equation}
Quantization
\begin{equation}
i\{ ..\hs{1mm} ,\hs{1mm}  ..\}\rightarrow [..\hs{1mm} ,\hs{1mm} ..]
\end{equation}
Commutator algebras
\begin{equation}
[J_{am},J_{bn}]=-if_{~ab}^{c}J_{c\hspace{.2mm}m+n}-\frac{k}{2}m\eta_{ab}
\delta_{m+n,0}
\end{equation}
\begin{equation}
[T_m,T_n]=(m-n)T_{m+n}+{c\over 12}(m^3-m)\delta_{m,-n}
\end{equation}
\begin{equation}
(J_{am})^\dagger=J_{a-m}, \hs{5mm} a=0,1,2, \hs{5mm} (T_m)^\dagger=T_{-m}
\end{equation}
New basis:
\begin{equation}
J_{\pm m}=\frac{1}{\sqrt{2}}(iJ_{1m}\mp J_{2m}), \hsv (J_{\pm m})^\dagger=-J_{\mp -m}, \hs{5mm}
J_{3 m}^\dagger=J_{3 -m}
\end{equation}
\begin{equation}
[J^{+}_{m},J^{-}_{m}]=J^{0}_{m+n}+\frac{k}{2}m\delta_{m+n},\hs{2mm} [J^{0}_{m},J^{\pm}_{n}]=\pm J^{\pm}_{m+n},\hs{2mm} [J^{0}_{m},J^{0}_{n}]=\frac{k}{2}m\delta_{m+n}\vs{10mm}
\end{equation}
Virasoro generators
\begin{equation}
L_{n} = \frac{1}{k+2}\sum_{m\in\mathbb{Z}}:\left[J^{+}_{m}J^{-}_{n-m} + J^{-}_{m}J^{+}_{n-m} + J^{0}_{m}J^{0}_{n-m}\right]:
\end{equation}
Connection to Einstein-Hilbert action
\begin{equation}
I_{EH}=\frac{1}{16\pi G}\int_{\cal{M}}d^3x\sqrt{-g}(R-2\La)=I_{CS}[A^+]-I_{CS}[A^-]
\end{equation}
where 
\begin{equation}\La=-\frac{~1~}{l^2}\end{equation} 
\begin{equation}k=-\frac{l}{4G}\end{equation} and
\begin{equation}A^\pm=\omega\pm\frac{1}{l}e=(\omega^a\pm\frac{1}{l}e^a)T_a
\end{equation}
Conventions for gravity
\begin{equation}
de^a+\omega^a_{~b}\wedge e^b=0
\end{equation}
\begin{equation}
\omega^a={1\over 2}\ep^{abc}\omega_{bc},\hsv \omega_{ab}=\eta_{ac}\omega^c_{~b}=-\omega
_{ba},\hsv \omega^{ab}=-\ep^{abc}\omega_c\end{equation}
\begin{equation}
R^a_{~b}=\half R^a_{~bcd}e^c\wedge e^d=\half R^a_{~b\mu\nu}dx^\mu\wedge dx^\nu =d\omega^a_{~b}+\omega^a_{~c}\wedge\omega^c_{~b}
\end{equation}
Ricci tensor and the curvature scalar:
\begin{equation}
{\cal R}^a_{~b}=R^{ac}_{~~bc}\hsv {\cal R}^\mu_{~\nu}=R^{\mu\alpha}_{~~\nu\alpha}\hsv 
{\cal R}={\cal R}^a_{~a}=R^{ac}_{~~ac}=R^{\mu\nu}_{~~\mu\nu}
\end{equation}
Einstein's equations (from the action above):
\begin{equation}
{\cal R}_{\al\beta}-\half{\cal R}g_{\al\beta}+\La g_{\al\beta}=0\end{equation}


\newpage

\begin{thebibliography}{AA}
\bibitem{BTZ} M. Banados, C. Teitelboim and J. Zanelli, {\em The black hole in three-dimensional spacetime}, Phys. Rev. Lett. {\bf 69} (1992) 1849-1851, {\tt hep-th/9204099}
\bibitem{Carlip98} S. Carlip, {\em What we don't know about BTZ black hole entropy}, Class. Quant. Grav. {\bf 15} (1998) 3609-3625, {\tt hep-th/9806026}
\bibitem{StrVafa} A. Strominger, C. Vafa, {\em Microscopic origin of the Bekenstein-Hawking entropy}, Phys. Lett. {\bf B379} (1996) 99-104, {\tt hep-th/9601029}
\bibitem{Bekenstein} J.D. Bekenstein, {\em Black holes and entropy}, Phys. Rev. {\bf D9} (1973) 2333-2346
\bibitem{Hawking} S. Hawking, {\em Gravitational radiation from black holes}, Phys.Rev.Lett. 26 (1971) 1344-1346
\bibitem{Hwang:modinv} M. Henningson, S. Hwang, P. Roberts, B. Sundborg, {\em Modular invariance of $SU(1,1)$ strings}, Phys. Lett. {\bf 267B} (1991) 350-355
\bibitem{Hwang:modinv2} S. Hwang and P. Roberts, {\em Interaction and modular invariance of strings on curved manifolds}, Proceedings of the 16th Johns Hopkins Workshop, "Current Problems in Particle Theory" (G\"{o}teborg, Sweden, 1992) ed: L. Brink and R. Marnelius, {\tt hep-th/9211075}
\bibitem{Maldacena:wzw1} J. Maldacena and H. Ooguri, {\em Strings in AdS$_{3}$ and the $SL(2,\RR)$ WZW model. Part 1: The Spectrum},  (2000) {\tt hep-th/0001053}
\bibitem{Strominger98} A. Strominger, {\em Black hole entropy from near-horizon microstates}, JHEP {\bf 9802} (1998) 009, {\tt hep-th/9712251}
\bibitem{Brown-Henneaux} J. D. Brown and M. Henneaux, {\em Central charges in the canonical realization of asymptotic symmetries: an example from three-dimensional gravity}, Commun. Math. Phys. {\bf 104} (1986) 207-226
\bibitem{Fjelstad-Hwang} J. Fjelstad and S. Hwang, {\em Equivalence of Chern-Simons gauge theory and WZNW model using a BRST symmetry}, Phys. Lett. {\bf B466} (1999) 227-233, {\tt hep-th/9906123}
\bibitem{Townsend:csgrav} A. Achucarro and P. K. Townsend, {\em A Chern-Simons action for three-dimensional anti-de Sitter supergravity theories}, Phys. Lett. {\bf B180} (1986) 89
\bibitem{Witten:gravcs} E. Witten, {\em (2+1)-dimensional gravity as an exactly soluble system}, Nucl. Phys. {\bf B311} (1988) 46
\bibitem{Banados95} M. Banados, {\em Global charges in Chern-Simons theory and the 2+1 black hole}, Phys. Rev. {\bf D52} (1995) 5816 {\tt hep-th/9405171}
\bibitem{Carlip95} S. Carlip, {\em The statistical mechanics of the (2+1)-dimensional black hole}, Phys. Rev. {\bf D51} (1995) 632-637, {\tt gr-qc/9409052}
\bibitem{Schonfeld} J. F. Schonfeld, {\em A mass term for three-dimensional gauge fields}, Nucl. Phys. {\bf B185} (1981) 157-171
\bibitem{JackTemp} R. Jackiw and S. Templeton, {\em How super-renormalizable interactions cure their infrared divergences}, Phys. Rev. {\bf D23} (1981) 2291
\bibitem{DesJackTemp1} S. Deser, R. Jackiw and S. Templeton, {\em Three-dimensional massive gauge theories}, Phys. Rev. Lett. {\bf 48} (1982) 975
\bibitem{DesJackTemp2} S. Deser, R. Jackiw and S. Templeton, {\em Topologically massive gauge theories}, Ann. Phys {\bf 140} (1982) 372-411
\bibitem{Sundborg:multiadsbh} T. M{\aa}nsson and B. Sundborg, {\em Multi-black hole sectors of AdS$_{3}$ gravity}, (2000) {\tt hep-th/0010083}
\bibitem{Witten:qftjones} E. Witten, {\em Quantum field theory and the Jones polynomial}, Commun. Math. Phys. {\bf 121} (1989) 351
\bibitem{Fradkin:bfv1} E. S. Fradkin and G. A. Vilkovisky, {\em Quantization of relativistic systems with constraints}, Phys. Lett. {\bf B55} (1975) 224
\bibitem{Batalin:bfv2} I. A. Batalin and G. A. Vilkovisky, {\em Relativistic S matrix of dynamical systems with boson and fermion constraints}, Phys. Lett. {\bf B69} (1977) 309-312
\bibitem{Hwang91} S. Hwang, {\em No ghost theorem for SU(1,1) string theories}, Nucl. Phys. {\bf B354} (1991) 100-112 
\bibitem{Hwang92} S. Hwang, {\em Cosets as gauge slices in SU(1,1) strings}, Phys. Lett. {\bf B276} (1992) 451-454, {\tt hep-th/9110039 }
\bibitem{DLP} L.J. Dixon, J. Lykken and M.E. Peskin, {\em N=2 superconformal symmetry and SO(2,1) current algebra}, Nucl. Phys. {\bf B325} (1989) 329-355 
\bibitem{Carlip:bhreview} S. Carlip, {\em The (2+1)-dimensional black hole}, Class. Quant. Grav. {\bf 12} (1995) 2853-2879 {\tt gr-qc/9506079}
\bibitem{Maldacena-Maoz} J. Maldacena and L. Maoz, {\em De-singularization by rotation}, {\tt hep-th/0012025}
\bibitem{Dabholkar1} A. Dabholkar, {\em Strings on a cone and black hole entropy}, Nucl. Phys. {\bf B439} (1995) 650-664 {\tt hep-th/9408098}
\bibitem{Dabholkar2} A. Dabholkar, {\em Quantum corrections to black hole entropy}, Phys. Lett. {\bf B347} (1995) 222-229 {\tt hep-th/9409158}
\bibitem{MartMcElg} E. Martinec and W. McElgin, {\em String theory on AdS orbifolds}, (2001) {\tt hep-th/0106171}
\bibitem{Son} J. Son, {\em String theory on \ads$/\mathbb{Z}_N$}, (2001) {\tt hep-th/0107131}
\bibitem{KeskHem} S. Hemming and E. Keski-Vakkuri, {\em The spectrum of strings on BTZ black holes and spectral flow in the SL(2,R) WZW model}, (2001) {\tt hep-th/0110252}
\bibitem{Cangemi-L-M}D. Cangemi, M. Leblanc,
R.B. Mann, {\em Gauge formulation of the spinning black hole in (2+1) dimensional anti deSitter space}, Phys. Rev {\bf D48} (1993) 3606-3610, {\tt gr-qc/9211013}
\bibitem{Banados-H-T-Z}M. Banados, M. Henneaux, C. Teitelboim and J. Zanelli, {\em Geometry of the 2+1 dimensional black hole}, Phys. Rev {\bf D48} (1993) 1506-1525, {\tt gr-qc/9302012}
\bibitem{Matschull} H.-J. Matschull, {\em On the relation between 2+1 Einstein gravity and Chern Simons theory}, Class. Quant. Grav. {\bf 16} (1999) 2599-2609, {\tt gr-qc/9903040}
\bibitem{Strominger:dSCFT} A. Strominger, {\em The dS/CFT correspondence}, (2001), {\tt hep-th/0106113}
\bibitem{DesJack} S. Deser and R. Jackiw, {\em String sources in $2+1$-dimensional gravity}, Ann. Phys. {\bf 192} (1989) 352-367
\bibitem{blackstring1} J. H. Horne, G. T. Horowitz, {\em Exact black string solutions in three dimensions}, Nucl. Phys {\bf B368} (1992) 444-462, {\tt hep-th/9108001}
\bibitem{blackstring2} N.S. Deger, A. Kaya, E. Sezgin, P. Sundell, {\em Matter coupled \ads supergravities and their black strings}, Nucl. Phys. {\bf B573} (2000) 275-290, {\tt hep-th/9908089}

\end{thebibliography}
\end{document}